\documentclass[fleqn,usenatbib]{mnras}

\usepackage[T1]{fontenc}

\DeclareRobustCommand{\VAN}[3]{#2}
\let\VANthebibliography\thebibliography
\def\thebibliography{\DeclareRobustCommand{\VAN}[3]{##3}\VANthebibliography}

\usepackage{graphicx}	
\graphicspath{ {./figs/} }
\usepackage{amsmath}	
\usepackage{amssymb}	
\usepackage{physics}
\usepackage{soul} 

\newcommand*\mean[1]{\overline{#1}}
\newcommand*\avg[1]{\left\langle{#1}\right\rangle}
\newcommand{\ra}{Ra}
\newcommand{\rac}{Ra_\mathrm{c}}
\newcommand{\ek}{E}
\newcommand{\pr}{Pr}
\newcommand{\prmag}{Pm}
\newcommand{\di}{Di}
\newcommand{\Ro}{Ro_\ell}

\newcommand{\fdip}{f_\mathrm{dip}}
\newcommand{\Nrho}{N_\rho}
\newcommand{\rin}{r_\mathrm{i}}
\newcommand{\rout}{r_\mathrm{o}}
\newcommand{\dcz}{D_\mathrm{cz}}

\newcommand{\ez}{\vu{e}_\mathrm{z}}

\newcommand{\Cwz}{\mathcal{C}_{\omega\mathrm{z}}}
\newcommand{\rHz}{{\mathcal{H}^{\mathrm{rel}}_{\mathrm{z}}}}

\newcommand{\Fcor}{\mathcal{F}_{C}}
\newcommand{\Fpre}{\mathcal{F}_{P}}
\newcommand{\Flor}{\mathcal{F}_{L}}
\newcommand{\Finer}{\mathcal{F}_{I}}
\newcommand{\Fbuo}{\mathcal{F}_{B}}
\newcommand{\Fvis}{\mathcal{F}_{V}}

\def\vec#1{\ensuremath{\mathchoice{\mbox{\boldmath$\displaystyle#1$}}
{\mbox{\boldmath$\textstyle#1$}}
{\mbox{\boldmath$\scriptstyle#1$}}
{\mbox{\boldmath$\scriptscriptstyle#1$}}}}

\usepackage{newtxtext,newtxmath}

\title[Transition from multipolar to dipolar dynamos]{Transition from multipolar to dipolar dynamos in stratified systems}

\author[B. Zaire et al.]{
B. Zaire$^{1,2}$\thanks{E-mail: zaire@fisica.ufmg.br},
L. Jouve$^{1}$,
T. Gastine$^{3}$,
J-F. Donati$^{1}$,
J. Morin$^{4}$,
N. Landin$^{5}$,
and
C. P. Folsom$^{6}$
\\
$^{1}$IRAP, Université de Toulouse, CNRS / UMR 5277, CNES, UPS, 14 avenue E. Belin, Toulouse, F-31400 France\\
$^{2}$Universidade Federal de Minas Gerais, Belo Horizonte, MG, 31270-901, Brazil \\
$^{3}$Université de Paris, Institut de Physique du Globe de Paris,  UMR 7154 CNRS, 1 rue Jussieu, F-75005 Paris, France \\
$^{4}$LUPM, Université de Montpellier, CNRS, Place Eugène Bataillon, F-34095 Montpellier, France \\
$^{5}$Universidade Federal de Viçosa, Campus UFV Florestal, Florestal, 35690-000, MG, Brazil \\
$^{6}$Tartu Observatory, University of Tartu, Observatooriumi 1, Tõravere, 61602 Tartumaa, Estonia
}

\date{Accepted XXX. Received YYY; in original form ZZZ}

\pubyear{2022}

\begin{document}
\label{firstpage}
\pagerange{\pageref{firstpage}--\pageref{lastpage}}
\maketitle

\begin{abstract} Observations of surface magnetic fields of cool stars reveal a large diversity of configurations. Although there is now a consensus that these fields are generated through dynamo processes occurring within the convective zone, the physical mechanism driving such a variety of field topologies is still debated. This paper discusses the possible origins of dipole and multipole-dominated morphologies using three-dimensional numerical simulations of stratified systems where the magnetic feedback on the fluid motion is significant. Our main result is that dipolar solutions are found at Rossby numbers up to 0.4 in strongly stratified simulations, where previous works suggested that only multipolar fields should exist. We argue that these simulations are reminiscent of the outlier stars observed at Rossby numbers larger than 0.1, whose large-scale magnetic field is dominated by their axisymmetric poloidal component. As suggested in previous Boussinesq calculations, the relative importance of inertial over Lorentz forces is again controlling the dipolar to multipolar transition.  Alternatively, we find that the ratio of kinetic to magnetic energies can equally well capture the transition in the field morphology. We test the ability of this new proxy to predict the magnetic morphology of a few M-dwarf stars whose internal structure matches that of our simulations and for which homogeneous magnetic field characterization is available. Finally, the magnitude of the differential rotation obtained in our simulations is compared to actual measurements reported in the literature for M-dwarfs. In our simulations, we find a clear relationship between anti-solar differential rotation and the emergence of dipolar fields. 
\end{abstract}




\section{Introduction}
Over the last decade, spectropolarimetric observations coupled to tomographic inversion techniques enabled the reconstruction of the large-scale magnetic topology that stars host at their surfaces. Cool stars with significant convective envelopes (with spectral types later than G0) revealed a large diversity of magnetic morphologies \citep{DMP08,MDP10,FPB16,FBP18}. Fully convective stars typically are found to harbour strong poloidal fields with a significant dipolar component, while partly convective stars host more complex magnetic topologies, consisting of non-axisymmetric multipolar poloidal fields and significant toroidal fields \citep{DL09}. Although there is now a consensus that the magnetism of cool stars are generated through dynamo processes occurring within the outer convective zones \citep[see][for a recent review on the subject]{BB17}, the physical mechanism driving such a variety of large-scale field topologies is still debated.

The fact that both rotation and convection play a major role in the stellar dynamo process is, however, well established \citep[see e.g. activity proxy studies of][]{MP84,NHB84,PMM03,WND11,WNW18}. Their joint effect on the magnetic field generation becomes obvious when considering observational measurements of the large-scale fields of low-mass stars as a function of the non-dimensional Rossby number (defined as the ratio of inertial to Coriolis forces and traditionally computed as $Ro = P_\mathrm{rot}/\tau$, where $\tau$ is the convective turnover time and $P_\mathrm{rot}$ is the rotation period of the star). The averaged surface field strength $\langle B\rangle$ shows two clear trends with the Rossby number. For $Ro > 0.1$, spectropolarimetric observations show that the large-scale magnetic field of cool stars weakens with increasing Rossby number \citep{VGJ14,FPB16}. This parameter region is often called "the unsaturated regime" and follows $\langle B\rangle \propto Ro^{-1.40 \pm 0.10}$ \citep{SMF19}, where the toroidal component of the large-scale field is reported to weaken faster than the poloidal component \citep{PDS08,SJV15}. As the Rossby number decreases below the $Ro \sim 0.1$ threshold, cool stars enter the "saturated regime" in which the large-scale field strength is roughly constant \citep{DMP08}.

The Rossby number has also proved to be quite successful at distinguishing various magnetic field morphologies in stellar observations \citep{MDP10,FBP18}. Stars with masses lower than $0.5$\,M$_{\sun}$ and $Ro \lesssim 0.1$ happen to have simple (dipole dominated) surface magnetic fields, whereas most stars featuring more complex surface fields tend to have larger Rossby numbers. Based on these observational results, it has been argued that stellar magnetic fields increase in complexity for stars with higher Rossby numbers. However, counterexamples that include stars harbouring complex field structures at low $Ro$ and others hosting dipole-dominated magnetic morphologies at large $Ro$ \citep[with Rossby numbers ranging from 0.2 to 0.3 --][]{DMP08,FPB16,FBP18} question the idea of magnetic fields getting more complex for stars with higher Rossby numbers. These results indicate that although the Rossby number may help at distinguishing between various generation mechanisms for the stellar magnetic fields, other proxies need to be invoked to clearly understand the transition between dipole-dominated and more complex field structures.

In the last two decades, numerical simulations mimicking the interior of planets (and, to a lesser extent, stars) have focused on understanding the origins of the magnetic morphology produced by convective dynamos. Parametric studies were conducted, using the relative strength of the axial dipole as a topological diagnostic to characterize the large-scale magnetic field. Geodynamo simulations with a constant density across the convective zone \citep[e.g.,][]{CA06,OC06,SJ06,SKA12} advocated that the Rossby number is indeed a key factor regulating the magnetic morphology. These initial numerical experiments suggested that dipole dominated morphologies only occur when $Ro \lesssim 0.1$ (commonly referred to as "the dipolar branch"), while complex surface fields could exist at both low and high Rossby numbers. Nevertheless, very recently \citet{MPG20} and \citet{TGF21} performed geodynamo simulations to explore the influence of the Lorentz force on the dipole breakdown. The authors found that strong dipoles can be recovered at high-Rossby numbers (up to $Ro = 0.18$) provided that a significant Lorentz force is acting on the fluid, challenging the canonical use of the Rossby number to distinguish between dipolar and multipolar field geometries. They suggested the ratio of inertial over Lorentz forces as an alternative proxy to capture the dipolar-multipolar transition. We propose to test this appealing hypothesis when the effect of a density contrast is introduced in the system.

Similar to what was initially found in geodynamo studies, stellar dynamo simulations showed a dipolar-multipolar transition with the Rossby number when considering weak density contrasts \citep{GDW12,J14}. However, these studies found that the dipolar branch disappeared for increasing density contrast. The apparent disagreement between the magnetic morphology observed in stars and those obtained in simulations of stratified flows raised the important question of why numerical experiments were apparently preventing dipoles from existing when the density contrast is more realistic \citep{PR19}. Further explorations of stratified flows with different physical properties showed that dipoles could be recovered at $Ro \lesssim 0.1$ when modifying the relative importance of the forces acting on the flow \citep{SPR14,RPD15}. To our knowledge, the simulation of \citet{YCM15} with $Ro = 0.04$ corresponds to the highest density contrast in which dipolar dynamos are reported to date. The authors obtained a strong dipole after considering a reduced influence of the inertial force by adopting a high ratio of viscous to thermal diffusions in a simulation with a density contrast of $N_\rho = \ln{\rho_i/\rho_o} = 5$ (where $\rho_i$ and $\rho_o$ are the density at the bottom and top of the convective zone, respectively). These various numerical experiments suggest that the dipole collapse could be an artificial bias of the parameter space explored with simulations. Thus, a close look at the force balance is needed to assess if the chosen parameter regime is indeed relevant for stars.

In this work, we attempt at reproducing for the first time the dipole-dominated field morphologies observed in some stars with $Ro > 0.1$. To do so, we perform a systematic parametric study of 3D convective dynamo simulations with different Rossby numbers and density contrasts, both of which are important ingredients in the stellar dynamo context. Guided by previous geodynamo studies, we focus on regime where the Lorentz force is dynamically active on the flow. The paper is organized as follows: we discuss our dynamo model and the selected control parameters in Sec.~\ref{sec:dynmodel}. The magnetic field morphology obtained in our simulations is presented in Sec.~\ref{sec:magmorph}, while the physical mechanisms controlling it are explored in Sec.~\ref{sec:diptran}. In Sec.~\ref{sec:dyngen}, we examine more closely the magnetic field generation in our simulations. Finally, we compare our results with previous stellar and geodynamo simulations and explore their implications in light of stellar observations in Sec.~\ref{sec:conclusions}.
\section{Dynamo model} \label{sec:dynmodel}
\subsection{Governing equations} \label{gov_eqs}
We model a stratified fluid in a spherical shell with inner radius $\rin$ and outer radius $\rout$ that rotates with angular velocity $\Omega_o$ about the axis $\ez$. We solve the non-dimensional magneto-hydrodynamics (MHD) equations under the anelastic formulation of \citet{BR95} and \citet{LF99}, expressed by
\begin{equation} \label{eq:mom_conserv}
  \begin{split}
    \ek\left[\pdv{\va{u}}{t}  + (\va{u} \cdot \grad) \va{u} \right] &+ 2\vu{e}_z\cross\va{u} =  - \grad(\frac{p'}{\tilde{\rho}}) + \frac{\ra\ek}{\pr} gs'\vu{e}_\mathrm{r}  \\  
    &+ \frac{1}{\prmag \tilde{\rho}}(\curl \va{B} )\cross\va{B} + \frac{\ek}{\tilde{\rho}}\div S \qq*{,}
  \end{split} 
\end{equation}

\begin{equation} \label{eq:induction}
  \begin{split}
    \pdv{\va{B}}{t}   = \curl( \va{u}\cross\va{B} ) -\frac{1}{\prmag}\curl(\curl\va{B}) \qq*{,}  
  \end{split} 
\end{equation}

\begin{equation} \label{eq:energy_conserv}
  \begin{split}
    \tilde{\rho}\tilde{T}\left[\pdv{s'}{t}  + (\va{u} \cdot \grad) s' +u_\mathrm{r}\dv{\tilde{s}}{r} \right] = &\frac{1}{\pr}  \div(\tilde{\rho} \tilde{T} \grad s') + \frac{\pr\di}{\ra}Q_\nu \\
    &+ \frac{\pr\di}{\prmag^2\ek\ra}(\curl\va{B})^2 ,
  \end{split} \\
\end{equation}

\begin{eqnarray}   
    \div(\tilde{\rho} \va{u}) &=& 0 \qq*{,}  \\
    \div \va{B} &=& 0, \label{eq:div}
\end{eqnarray}
where $\va{u}$ is the velocity field, $\va{B}$ is the magnetic field,  $S$ represents the strain-rate tensor given by $$ S = \pdv{x_j} \left[\tilde{\rho}\left( \pdv{u_i}{x_j} + \pdv{u_j}{x_i}\right)\right]  - \frac{2}{3} \pdv{x_i} \left(\tilde{\rho}\pdv{u_j}{x_j}\right),
$$ and $Q_\nu$ is the viscous heating expressed as $$ Q_\nu = \tilde{\rho}\left( \pdv{u_i}{x_j} + \pdv{u_j}{x_i}  - \frac{2}{3} \delta_{ij} \div \va{u}\right) \pdv{u_i}{x_j}.
$$ Pressure and entropy fluctuations ($p'$ and $s'$, respectively) are defined with respect to the reference state (see Subsec.~\ref{sec:refstate}). We adopt a dimensionless formulation where the reference length scale is $\rout$ and the time is given in units of $\tau_\nu = \rout^2/\nu$, where $\nu$ is the fluid viscosity. The entropy scale is set to $\rout|\mathrm{d}{\tilde{s}}/\mathrm{d}{r}|_{r_o}$, where $|\mathrm{d}{\tilde{s}}/\mathrm{d}{r}|_{r_o}$ is the normalized background entropy gradient at the outer boundary  (see Sec.~\ref{sec:refstate}). The magnetic field is given in units of $\sqrt{\rho_{o}\mu\lambda\Omega_o}$, where $\mu$ is the magnetic permeability and $\lambda$ is the magnetic diffusivity. The gravity, density, and temperature are normalised by their outer radius values given by $g_{o}$, $\rho_{o}$, and $T_{o}$, respectively.

The dimensionless control parameters that appear in the equations above are the Ekman number $(\ek)$, Rayleigh number $(\ra)$, Prandtl number $(\pr)$, magnetic Prandtl number $(\prmag)$, and dissipation number $(\di)$. They are defined as
 \begin{displaymath} 
\ek = \frac{\nu}{\Omega_o \rout^2}\, , \;
\ra = \frac{g_{o} \rout^4}{c_\mathrm{p}\kappa\nu} \left| \dv{\tilde{s}}{r} \right|_{\rout} \, , \;
\pr = \frac{\nu}{\kappa} \, , \;
\prmag = \frac{\nu}{\lambda} \, , \;
\di = \frac{g_{o} \rout}{c_\mathrm{p} T_{o}},
\end{displaymath}
where $\kappa$ is the thermal diffusivity and $c_\mathrm{p}$ is the specific heat at constant pressure. We note that in the anelastic formulation adopted here, a non-adiabatic reference state is used. This translates into the appearance of a non-zero background entropy gradient $\dv{\tilde{s}}{r}$ in the entropy equation (Eq.~\ref{eq:energy_conserv}). The details of this reference state are discussed below.

\subsection{Reference state} \label{sec:refstate}
Thermodynamical quantities in Eqs.~\ref{eq:mom_conserv} to \ref{eq:energy_conserv} are expressed in terms of a reference (static) state and fluctuations around it. We adopt as reference state a nearly adiabatic ideal gas for which we prescribe the background entropy gradient $\dv{\tilde{s}}{r}$. We then deduce the reference temperature and density by solving the following equations: 
\begin{equation} 
\frac{1}{\tilde{T}}\pdv{\tilde{T}}{r}  =  \epsilon_\mathrm{s} \dv{\tilde{s}}{r}  - \frac{\di}{T_\mathrm{o}} g(r)
\end{equation}
and 
\begin{equation} 
\frac{1}{\tilde{\rho}}\pdv{\tilde{\rho}}{r}  =  \epsilon_\mathrm{s} \dv{\tilde{s}}{r}  - \frac{\di c_\mathrm{v}}{ (c_\mathrm{p} - c_\mathrm{v})T_\mathrm{o}}g(r),
\end{equation}
where we set the control parameter $\epsilon_\mathrm{s} = 10^{-4} \ll 1$, which is a necessary condition to ensure that we are still close to an adiabatic state. This formulation with a prescribed non-adiabaticity $\mathrm{d}\tilde{s}/\mathrm{d}r$ allows us to control the energy transport inside the star (notice its presence in Eq.~\ref{eq:energy_conserv}) and has been previously adopted in numerical models of gas giant planets \citep{DW18,GW21}. The background entropy sets radiative regions whenever $\mathrm{d}\tilde{s}/\mathrm{d}r > 0 $, while convectively-unstable regions occur when $ \mathrm{d}\tilde{s}/\mathrm{d}r < 0 $. 

\begin{figure}
     \centering
     \includegraphics[width=0.48\textwidth]{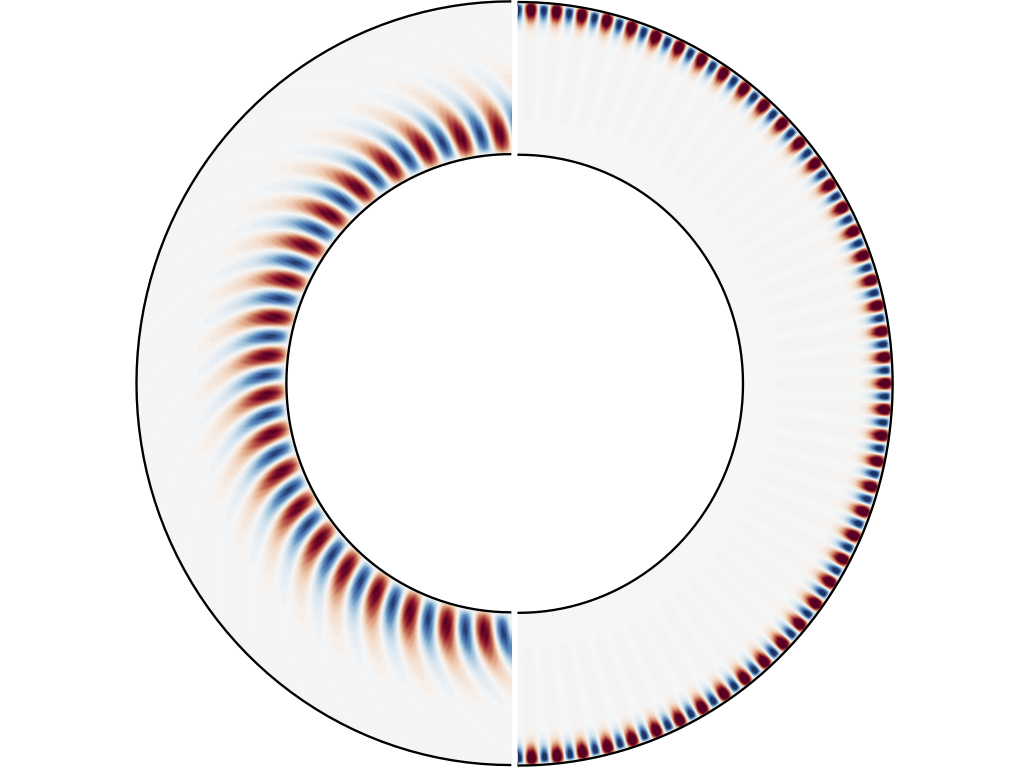}
     \caption{Structure of the most unstable mode for convection forced through our background entropy profile (left) and through a more traditional entropy profile (right) for a density contrast $N_\rho=3$. Represented on the figure is an equatorial cut of the radial velocity close to the onset of convection at the values of $\ek=1.6\times 10^{-5}$ and $\pr=1$.}
     \label{fig:unstable}
\end{figure}

In the present work, we simulate convective shells with $\rin/\rout = 0.6$ and a fixed background entropy gradient $\mathrm{d}\tilde{s}/\mathrm{d}r = - 1$. We note that this choice is motivated by the fact that the entropy gradient calculated from 1D stellar evolution models of Sun-like stars is indeed approximately constant in the bulk of the convection zone (i.e., excluding the outer $5\%$ of the star in radius), which is the region we aim at modelling in this work. Our background entropy profile thus differs from previous anelastic studies, like the ones presented in the anelastic benchmark of \citet{JBB11}, where the reference state entropy is the solution of a conduction equation on which conditions of fixed entropy are applied. This leads to a solution with a gradient varying with radius, the maximal values of which being located in the outer part of the spherical shell. In our case, the gradient is constant throughout the shell, leading to a more homogeneous forcing of convection. This difference is illustrated in Figure~\ref{fig:unstable}, where the structure of the most unstable mode at the onset of convection is shown for our present work (left) and for an adiabatic reference state as used in \citet{JKM09} (right) with the same values of $N_\rho$, $\ek$ and $\pr$. At onset, our forcing of convection results in unstable modes located close to the bottom boundary \citep[see also][for similar results with an adiabatic reference state but different boundary conditions]{CH18}. When the Rayleigh number is increased however, strong convective velocities build close to the outer shell, as expected in stratified systems. To be more specific, we now give in Table~\ref{tab:critical} the values of the critical Rayleigh number and the critical azimuthal wavenumber in our setup, determined numerically at the different density contrasts used in our simulations and for the values of $\ek$ and $\pr$ adopted in all our calculations and which are specified in the next Subsection~\ref{subsec:param}.

\begin{table}
    \caption{Critical Rayleigh numbers and azimuthal wavenumbers for our setup, for the three different density contrasts used in our simulations. These numbers are determined without taking into account the presence of a magnetic field.}
    \label{tab:critical}
    \centering
    \begin{tabular}{ccc}
    \hline
    $\Nrho$  & ${\mathrm{\rac}}$ & $m_c$ \\
    \hline
    1 & $1.92\times 10^7$& 32 \\
    1.5 & $2.40\times 10^7$& 37 \\
    3 & $3.56\times 10^7$& 39\\
    \hline
    \end{tabular}
\end{table}

We adopt a physically-motivated gravity based on the reference state of a main-sequence cool star that reads 
\begin{equation} \label{eq:grav}
    g(r) = -\frac{7.36~r}{r_\mathrm{o}} +  \frac{4.99~r^2}{r_\mathrm{o}^2} +  \frac{3.71~r_\mathrm{o}}{r} -  \frac{0.34~r_\mathrm{o}^2}{r^2}.
\end{equation}
For the radial domain explored in this paper (with radius ratio $\rin/\rout = 0.6$), this gravity profile is virtually identical to the point mass approximation used in many parametric studies investigating dynamo action in planets and stars. We expect thus that any differences between our simulations and other similar ones in the literature with $\rin/\rout = 0.6$ are most likely caused by differences in the background entropy profile or control parameters (see Sec.~\ref{subsec:param}) rather than in the gravity profile.

\subsection{Numerical model and boundary conditions}
We use the anelastic version of the open-source code \verb|MagIC| \citep[][freely available at \url{https://github.com/magic-sph/magic}]{GW12} to solve Eqs.~\ref{eq:mom_conserv} to \ref{eq:div} in spherical coordinates. \verb|MagIC| has been validated through several anelastic benchmarks \citep{JBB11}. To evolve the Eqs.~\ref{eq:mom_conserv}-\ref{eq:energy_conserv} in time a mixed algorithm is adopted, where linear terms (except for the Coriolis one) are treated implicitly and non-linear terms are handled explicitly. Spherical harmonics are used as basis functions of the angular coordinates $(\theta, \phi)$ and are handled using the \verb|SHTns| library \citep[][freely available at \url{https://bitbucket.org/nschaeff/shtns}]{S13}. These functions are truncated at a maximum degree $\ell_\mathrm{max}$, sufficient to capture physical processes at play (typically ranging from $213$ to $341$ in our simulations). Chebyshev polynomials are used in the radial direction along with the mapping proposed by \citet{KT93}, which alleviates the grid refinement created near inner and outer boundaries in the standard formulation of the Chebyshev-collocation points. We refer to \citet{GW21} for additional details of this implementation in \verb|MagIC|. 

In the full set of simulations, we adopt stress-free boundary conditions on the velocity field, 
\begin{equation}
    u_r = \pdv{r}(\frac{u_\theta}{r}) = \pdv{r}(\frac{u_\phi}{r}) = 0 \qq{on} r=\rin \qand r = \rout,
\end{equation}
potential field boundaries on the magnetic field,
\begin{equation}
    \va{J} = \curl\va{B} = 0 \qq{on} r=\rin \qand r = \rout,
\end{equation}
and fixed entropy values, set to 0, at both boundaries. We initialize the velocity field with a small-amplitude random perturbation. The initial magnetic field is set to a dipole of strength $\Lambda = 0.44$ at the bottom of the convective zone (i.e., at $r = \rin$), where $\Lambda = \mean{\expval{B^2}}$ is the Elsasser number expressed in terms of the dimensionless magnetic field.

\subsection{Choice of parameters}
\label{subsec:param}
In order to perform stellar dynamo simulations, a crucial ingredient to take into account is the density stratification. In the main-sequence, cool stars show a density contrast between the bottom ($\rho_i$) and the top ($\rho_o$) of the convective zone that can reach $N_\rho \sim 11$ \citep[according to models generated with the ATON code,][]{LVD06}. However, density contrasts as high as those seen in stars cannot be attained by numerical simulations as it drives fast small-scale motions that are too computationally demanding. In order to bypass this limitation, some authors chose to exclude from the numerical domain the outer few per cent of the stellar radii where the sharpest density gradients exist \citep{DSB06,B08,BMB11,ZGK16,EBS17,GZS19}. 
We here also exclude this sharp gradient region from our domain and study the effect of varying $N_\rho$ from 1 to 3 to assess the influence of an increase of the density contrast on the magnetic field generation and flow dynamics.

We consider three different setups with $N_\rho  = 1$, $1.5$, and $3$. These density contrasts are practically achieved in our formulation after fixing the dissipation number $\di = 1.53$, $2.7$, and $10$, respectively. Following previous studies, we adopt moderate values of $\ek = 1.6 \times 10^{-5}$ and $\pr = 1$ that reduce the numerical cost of each simulation, allowing us to perform a parametric study varying the Rayleigh number for the three different density contrasts. We increase the Rayleigh number from $1.3$ to $32.7~\rac$ to explore the implications of distinct turbulence levels on the magnetic field morphology, where the convective onset $\rac$ varies depending on the density contrast over the convective zone (see Table \ref{tab:critical}).

We are thus left with the choice of the magnetic Prandtl number $\prmag$. Recent studies \citep[e.g.,][]{D16,DOP18,SGA19} have advocated that pushing a single parameter closer to the values observed in astrophysical objects may not represent the correct force balance at stake \citep[e.g., $\ek \approx 10^{-13}$, $\pr \approx 10^{-7}$, and $\prmag \approx 10^{-3}$ at the bottom of the Solar convective zone;][]{O03}. There is considerable evidence from numerical simulations with/without density contrast that there is a critical magnetic Prandtl number $\prmag_c$ below which dipolar dynamo solutions cannot be achieved for a fixed Ekman number. This brings some concerns as strong dipoles are observed in stars \citep[e.g.,][]{DL09}. One potential way to overcome this limitation is to adopt $\prmag > \prmag_c$. However, previous works showed that $\prmag_c$ varies with $\ek$ and $N_\rho$. For the value adopted in this work of $\ek = 1.6 \times 10^{-5}$, it was shown that the critical magnetic Prandtl number obeys the relation $\prmag_c = 2 N_\rho - 2$ \citep{SPR14}. Therefore, we choose to fix $\prmag = 5$ for the entire set of simulations, which is greater than the critical value obtained for the highest stratified setup $N_\rho=3$. Moreover, we initialize our simulations with a dipole of strength $\Lambda = 0.44$, which has the same order of magnitude of typical stellar strengths \citep[e.g.,][]{MDP08,GMD13}.

\section{Results} 
We performed altogether 23 dynamo simulations with different density contrasts and Rayleigh numbers. We ran numerical models for a few magnetic diffusion times to achieve meaningful dynamo steady-states, which resulted here in rather costly simulations. The journal of simulations is summarized in Table~\ref{tab:runs}.
We provide the total simulation time $\tau_\text{end}$ in units of magnetic diffusion time, which we defined as
\begin{equation} \label{eq:taulambda}
    \tau_\lambda = \frac{\dcz^2}{\lambda} = \prmag \left(\frac{\dcz}{\rout}\right)^2  \tau_\nu
\end{equation}
using the convective shell size $\dcz = \rout - \rin$ as the relevant length scale. Throughout this work, we employ overbars $\mean{\; \cdot \;}$ to represent averages over time, brackets $\expval{\cdot}$ to represent volume averages, and $\expval{\cdot}_i$ to represent spatial averages in the direction $\vu{e}_\mathrm{i}$. Time averages are performed only after the solutions have reached a well-established steady-state and typically cover a few magnetic diffusion times (for more information see Appendix~\ref{appendix:dipolarity}).

\begin{table*}
    \caption{Journal of simulations. First column yields the run ID. Columns 2-6 indicate the parameters imposed in each simulation (see Sec.~\ref{subsec:param}). Column 7 and 8 show the total simulation time $\tau_\text{end}$ and the averaging time $\tau_\mathrm{avg}$ (both expressed in units of magnetic diffusion time $\tau_\lambda$ as defined in Eq.~\ref{eq:taulambda}), respectively. Column 9 gives the dominant scale of convection $\ell_\text{peak}$ (Eq.~\ref{eq:lpeak}). Column 10 displays the local Rossby number (Eq.~\ref{eq:rossby}). Column 11 shows the Inertia over Lorentz force ratio (see Sec.~\ref{sec:fb}) and column 12 the kinetic over magnetic energy ratio (Eq.~\ref{eq:ekem}). Column 13 shows the dipolarity computed using Eq.~\ref{eq:fdip}, whereas column 14 gives a variation of the dipolarity measure based on the total dipole $f_\mathrm{dip,Tot.}$ (see discussion in Sec.~\ref{sec:magmorph}). }
    \setlength{\tabcolsep}{3.2pt}
    \label{tab:runs}
    \centering
    \begin{tabular}{lccccccccccccc}
    \hline
Run ID & $\Nrho$ & $\rho_i/\rho_o$ & $\ra$ &  $\ra$/$\rac$ & ($N_r, N_\theta, N_\phi$) & $\tau_\text{end}$ & $\tau_\text{avg}$ & $\ell_\text{peak}$ & $\Ro$ & $\mathcal{F}_\mathrm{I}/\mathcal{F}_\mathrm{L}$ & $E_{K}/E_{M}$ & $\fdip$ & $f_\mathrm{dip,Tot.}$ \\ 

&  &  &  &   &  & ($\tau_\lambda$) & ($\tau_\lambda$) & & &  & &  &  \\ 
\hline
FC01 & 1.0 & 2.7 & $4.77 \times 10^7 $ & 2.5 & $(73 ,320,640)$ &  5.1 &  2.0 &  17  & $ 0.023 \pm 0.004$ & $ 0.03 \pm 0.25$ & $ 0.11 \pm 0.17$ & $ 0.84 \pm 0.06$ & $ 0.84 \pm 0.06$ \\ 
FC02 & 1.0 & 2.7 & $6.25 \times 10^7 $ & 3.3 & $(73 ,320,640)$ &  4.3 &  2.0 &  18  & $ 0.036 \pm 0.007$ & $ 0.05 \pm 0.39$ & $ 0.05 \pm 0.01$ & $ 0.87 \pm 0.02$ & $ 0.87 \pm 0.02$ \\ 
FC03 & 1.0 & 2.7 & $7.81 \times 10^7 $ & 4.1 & $(73 ,320,640)$ &  4.1 &  2.0 &  19  & $ 0.05 \pm 0.01$ & $ 0.07 \pm 0.30$ & $ 0.08 \pm 0.01$ & $ 0.87 \pm 0.02$ & $ 0.87 \pm 0.01$ \\ 
FC04 & 1.0 & 2.7 & $1.04 \times 10^8 $ & 5.5 & $(73 ,512,1024)$ &  4.1 &  2.0 &  22  & $ 0.09 \pm 0.02$ & $ 0.12 \pm 0.11$ & $ 0.15 \pm 0.02$ & $ 0.87 \pm 0.01$ & $ 0.87 \pm 0.01$ \\ 
FC05 & 1.0 & 2.7 & $1.25 \times 10^8 $ & 6.5 & $(73 ,512,1024)$ &  0.9 &  0.3 &  28  & $ 0.12 \pm 0.03$ & $ 0.16 \pm 0.12$ & $ 0.26 \pm 0.03$ & $ 0.87 \pm 0.02$ & $ 0.88 \pm 0.02$ \\ 
FC06 & 1.0 & 2.7 & $1.56 \times 10^8 $ & 8.2 & $(73 ,512,1024)$ &  4.6 &  1.5 &  24  & $ 0.18 \pm 0.05$ & $ 0.49 \pm 0.09$ & $ 1.05 \pm 0.08$ & $ 0.12 \pm 0.03$ & $ 0.13 \pm 0.03$ \\ 
FC07 & 1.0 & 2.7 & $3.12 \times 10^8 $ & 16.3 & $(73 ,512,1024)$ &  3.1 &  1.0 &  19  & $ 0.32 \pm 0.09$ & $ 0.57 \pm 0.09$ & $ 1.10 \pm 0.09$ & $ 0.11 \pm 0.03$ & $ 0.12 \pm 0.02$ \\ 
FC08 & 1.0 & 2.7 & $6.25 \times 10^8 $ & 32.7 & $(73 ,1024,2048)$ &  1.3 &  0.5 &  14  & $ 0.53 \pm 0.12$ & $ 0.58 \pm 0.08$ & $ 1.36 \pm 0.10$ & $ 0.12 \pm 0.03$ & $ 0.18 \pm 0.03$ \\ 
\hline
FC09 & 1.5 & 4.4 & $4.77 \times 10^7 $ & 2.0 & $(73 ,320,640)$ &  4.3 &  1.0 &  45  & $ 0.031 \pm 0.007$ & $ 0.25 \pm 0.13$ & $ 0.59 \pm 0.30$ & $ 0.71 \pm 0.06$ & $ 0.71 \pm 0.06$ \\ 
FC10 & 1.5 & 4.4 & $6.25 \times 10^7 $ & 2.6 & $(73 ,320,640)$ &  4.5 &  1.4 &  38  & $ 0.05 \pm 0.01$ & $ 0.19 \pm 0.19$ & $ 0.58 \pm 0.25$ & $ 0.62 \pm 0.04$ & $ 0.62 \pm 0.04$ \\ 
FC11 & 1.5 & 4.4 & $7.81 \times 10^7 $ & 3.3 & $(73 ,320,640)$ &  6.5 &  2.5 &  38  & $ 0.07 \pm 0.02$ & $ 0.33 \pm 0.13$ & $ 0.70 \pm 0.19$ & $ 0.44 \pm 0.11$ & $ 0.45 \pm 0.11$ \\ 
FC12 & 1.5 & 4.4 & $1.04 \times 10^8 $ & 4.3 & $(73 ,512,1024)$ &  4.9 &  1.9 &  35  & $ 0.11 \pm 0.04$ & $ 0.28 \pm 0.11$ & $ 0.59 \pm 0.08$ & $ 0.15 \pm 0.04$ & $ 0.45 \pm 0.04$ \\ 
FC13 & 1.5 & 4.4 & $1.56 \times 10^8 $ & 6.5 & $(73 ,512,1024)$ &  5.1 &  1.5 &  35  & $ 0.17 \pm 0.06$ & $ 0.34 \pm 0.11$ & $ 0.70 \pm 0.09$ & $ 0.46 \pm 0.14$ & $ 0.56 \pm 0.08$ \\ 
FC14 & 1.5 & 4.4 & $3.12 \times 10^8 $ & 13.0 & $(73 ,512,1024)$ &  3.8 &  1.0 &  25  & $ 0.31 \pm 0.12$ & $ 0.57 \pm 0.09$ & $ 1.05 \pm 0.09$ & $ 0.12 \pm 0.03$ & $ 0.14 \pm 0.03$ \\ 
FC15 & 1.5 & 4.4 & $6.25 \times 10^8 $ & 26.0 & $(73 ,512,1024)$ &  1.5 &  0.5 &  20  & $ 0.52 \pm 0.19$ & $ 0.64 \pm 0.08$ & $ 1.35 \pm 0.09$ & $ 0.13 \pm 0.03$ & $ 0.17 \pm 0.03$ \\ 
\hline
FC16 & 3.0 & 19.3 & $4.77 \times 10^7 $ & 1.3 & $(73 ,320,640)$ &  8.1 &  1.9 &  42  & $ 0.013 \pm 0.003$ & $ 1.78 \pm 0.54$ & $ 1.17 \pm 0.18$ & $ 0.04 \pm 0.02$ & $ 0.12 \pm 0.04$ \\ 
FC17 & 3.0 & 19.3 & $7.81 \times 10^7 $ & 2.2 & $(73 ,320,640)$ &  5.4 &  1.5 &  38  & $ 0.037 \pm 0.008$ & $ 0.33 \pm 0.19$ & $ 0.41 \pm 0.10$ & $ 0.63 \pm 0.03$ & $ 0.63 \pm 0.03$ \\ 
FC18 & 3.0 & 19.3 & $1.56 \times 10^8 $ & 4.4 & $(73 ,512,1024)$ &  6.0 &  1.4 &  36  & $ 0.11 \pm 0.05$ & $ 0.36 \pm 0.11$ & $ 0.54 \pm 0.05$ & $ 0.54 \pm 0.03$ & $ 0.55 \pm 0.03$ \\ 
FC19 & 3.0 & 19.3 & $2.08 \times 10^8 $ & 5.8 & $(73 ,512,1024)$ &  2.4 &  1.0 &  39  & $ 0.15 \pm 0.08$ & $ 0.35 \pm 0.10$ & $ 0.58 \pm 0.06$ & $ 0.53 \pm 0.03$ & $ 0.54 \pm 0.03$ \\ 
FC20 & 3.0 & 19.3 & $3.12 \times 10^8 $ & 8.8 & $(73 ,512,1024)$ &  4.1 &  1.3 &  34  & $ 0.21 \pm 0.13$ & $ 0.36 \pm 0.09$ & $ 0.52 \pm 0.04$ & $ 0.63 \pm 0.05$ & $ 0.63 \pm 0.05$ \\ 
FC21 & 3.0 & 19.3 & $6.25 \times 10^8 $ & 17.6 & $(73 ,512,1024)$ &  1.6 &  0.5 &  31  & $ 0.38 \pm 0.25$ & $ 0.47 \pm 0.07$ & $ 0.64 \pm 0.06$ & $ 0.75 \pm 0.03$ & $ 0.75 \pm 0.03$ \\ 
FC22 & 3.0 & 19.3 & $7.44 \times 10^8 $ & 20.9 & $(73 ,1024,2048)$ &  1.1 &  0.4 &  30  & $ 0.41 \pm 0.26$ & $ 0.45 \pm 0.05$ & $ 0.68 \pm 0.05$ & $ 0.77 \pm 0.02$ & $ 0.77 \pm 0.02$ \\ 
FC23 & 3.0 & 19.3 & $9.20 \times 10^8 $ & 25.8 & $(73 ,1024,2048)$ &  1.3 &  0.4 &  29  & $ 0.51 \pm 0.32$ & $ 0.70 \pm 0.07$ & $ 1.20 \pm 0.08$ & $ 0.23 \pm 0.05$ & $ 0.25 \pm 0.05$ \\ 
\hline
\end{tabular}
\end{table*}

\subsection{Magnetic morphology} \label{sec:magmorph}
Since the physical origin of the various magnetic field morphologies observed in cool stars is still debated, in this study we particularly focus on the field topology achieved in our simulations. Traditionally, the magnetic field morphology has been assessed by measuring the relative importance of the axial-dipole at the stellar surface. This quantity, named \textit{dipolarity}, is defined as the relative strength of the axial-dipole\footnote{A different definition of `dipolarity' based on the relative energy of the axial dipole also appears in the literature, in which the right-hand-side of Eq.~\ref{eq:fdip} is squared.} \citep{CA06}: 
\begin{equation} \label{eq:fdip}
\fdip = \mean{\sqrt{\frac{ \iint \va{B}^2_{\ell=1, m=0} (r =r_\mathrm{o}, \theta, \phi) \sin{\theta}\dd{\theta}\dd{\phi}}{\sum_{\ell=1}^{11} \sum_{m=0}^{\ell} \iint \va{B}^2_{\ell,m} (r =r_\mathrm{o}, \theta, \phi) \sin{\theta}\dd{\theta}\dd{\phi}}}}.
\end{equation} 
Here, the normalization factor corresponds to the square root of the total surface magnetic energy stored in the largest spatial scales, i.e. in modes with order $\ell < 12$. It thus matches the typical resolution achieved in the surface magnetic field reconstruction of stars other than the Sun \citep[e.g., ][]{DMP08,MDP10,FPB16,FBP18}. We recall the reader that toroidal fields vanish at the outer boundary because of our magnetic boundary condition (and, therefore, only poloidal fields contribute in Eq.~\ref{eq:fdip}). Following previous authors \citep[e.g.,][]{OLD14,MPG20,TGF21}, we define simulations with $\fdip \ge 0.5$ (or equivalently, with an axial-dipole containing $25\%$ of the magnetic energy stored at modes up to $\ell = 11$) as dipolar dynamos. Conversely, simulations in which $\fdip < 0.5$ are defined as ``multipolar'' dynamos. The dipolarity measurements are given in Table~\ref{tab:runs} along with an alternative estimate based on the total dipole $f_\mathrm{dip,Tot.}$ (i.e., including the equatorial dipole contribution in the summation at the numerator of Eq.~\ref{eq:fdip}). We note that none of our simulations would change their classification as dipolar or multipolar dynamos if considering a dipolarity based on the total dipole. We thus stick to the  dipolarity definition given by Eq.~\ref{eq:fdip} throughout this work. 
\begin{figure}
     \centering
     \includegraphics[width=\columnwidth]{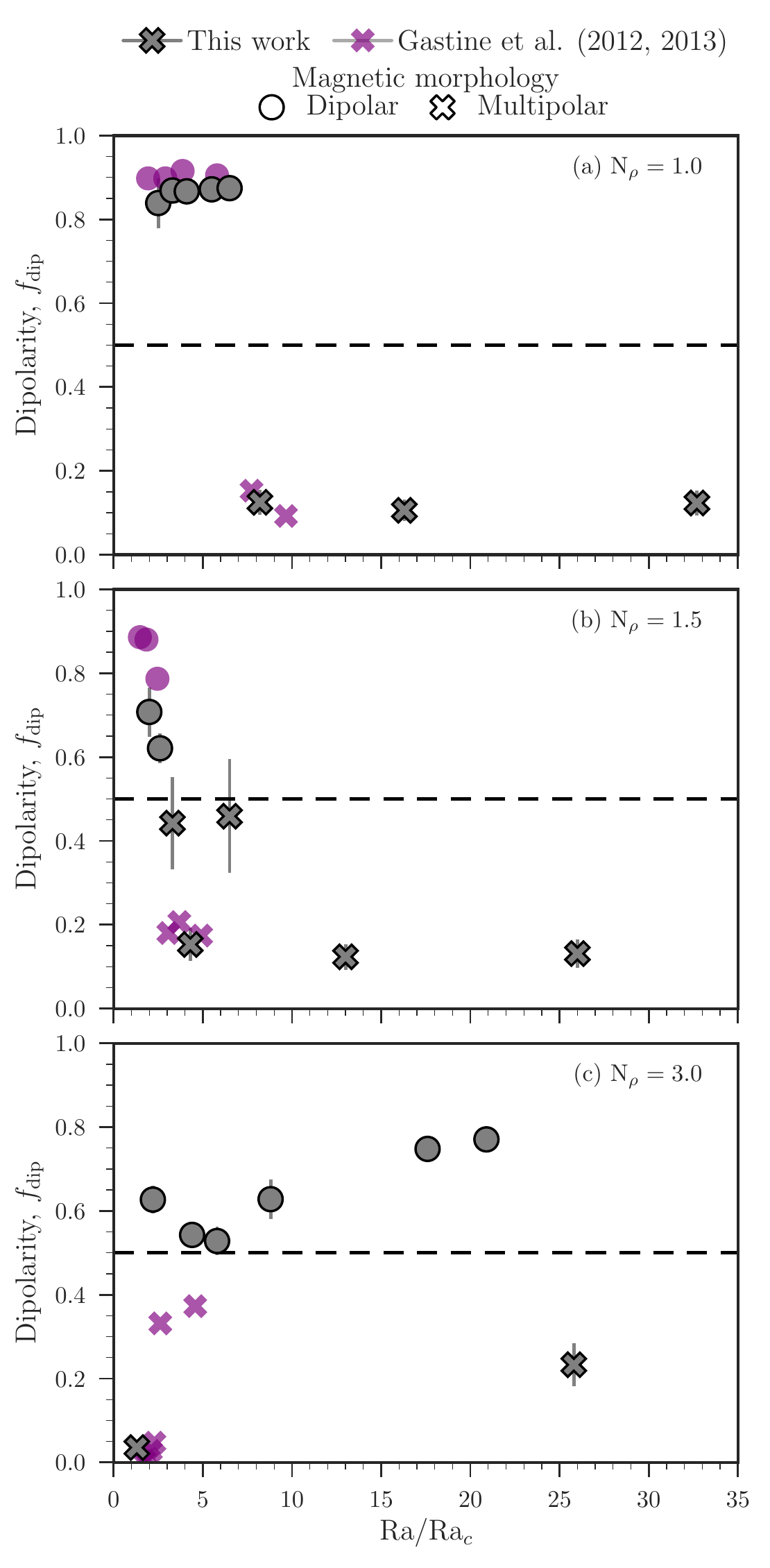}
     \caption{Surface dipolar fraction as a function of the Rayleigh number for the 23 runs listed in Table~\ref{tab:runs} (grey symbols). The shape of the symbols distinguishes between dipolar dynamos (circle) and multipolar dynamos (cross). Simulations with density contrast N$_\rho = \log{\rho_i/\rho_o} = 1$, $1.5$, and $3$, are separated respectively in panels (a), (b), and (c). Error bars represent one standard deviation about the time averaged dipolarity. Stratified dynamos with the same radius ratio ($\rin/\rout = 0.6$) and density contrasts, but $\prmag=1$ are included for comparison \citep[purple symbols;][]{GDW12,GMD13}.  
     }
     \label{fig:fdipRa}
\end{figure}

Figure~\ref{fig:fdipRa} shows how the dipolarity varies with the Rayleigh number. This figure shows three panels with $\fdip$ as a function of $\ra$, each at a particular $\Nrho$. Starting from the set of simulations with $\Nrho = 1$ (Figure~\ref{fig:fdipRa} a), we identify dipolar dynamos at low Rayleigh numbers followed by a sharp transition to multipolar dynamos as $\ra$ increases. This finding is in line with earlier simulations of \citet{GDW12,GMD13}\footnote{The control parameters adopted by \citet{GDW12,GMD13} coincide with those employed in this work with the exception of $\prmag$. However, with also different formulations of convective forcing (similar to what has been described in Figure~\ref{fig:unstable}), caution must be applied when attributing possible differences between the models to $\prmag$.} using $\prmag =1$ (purple symbols), which showed that the morphology transitions to a more complex configuration around $\ra = 7 \rac$. It also extends Rayleigh's parameter space coverage by about a factor of three when compared to \citet{GDW12,GMD13}, corroborating the hypothesis that only multipolar dynamos exist for forcings above the threshold leading to the dipole collapse  (i.e., $\ra \gtrsim 7 \rac$ for $\Nrho = 1$). 

The dipolarity trend, however, changes for the models with $N_\rho = 1.5$ (Figure~\ref{fig:fdipRa} b). While the \textit{plateau} with strong dipolar dynamos seen for the runs with $N_\rho = 1$ no longer exists, intermediate values of $\fdip$ appear, defining a rather continuous transition to the multipolar branch. We highlight that two of our multipolar cases are compatible with a dipole within error bars (estimated as one standard deviation over the time averaged value). An inspection of the simulations around $5~\rac$ reveals one case with polarity reversals (FC11) and two with excursions (FC12 and FC13) of the dipole field, thus explaining why large error bars are found in those cases where the dipolar field strongly varies in time. This finding is in accordance with previous studies evaluating reversing dipoles, which observed a tendency for its occurrence at Rayleigh numbers close to the transition between dipolar and multipolar dynamos \citep{KC02,OC06,WT10}. 

\begin{figure}
     \centering
     \includegraphics[width=\columnwidth]{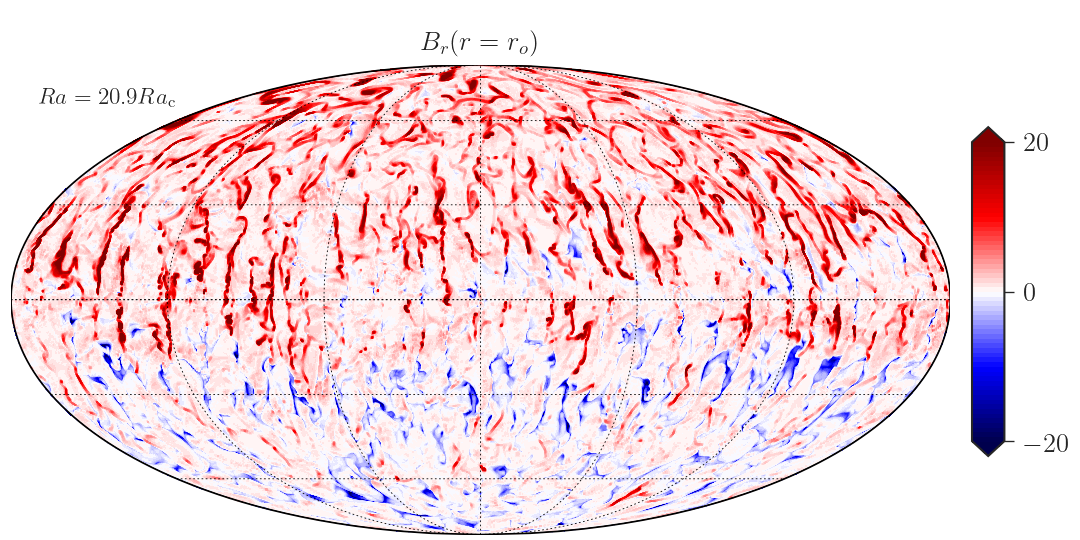}
     \includegraphics[width=\columnwidth]{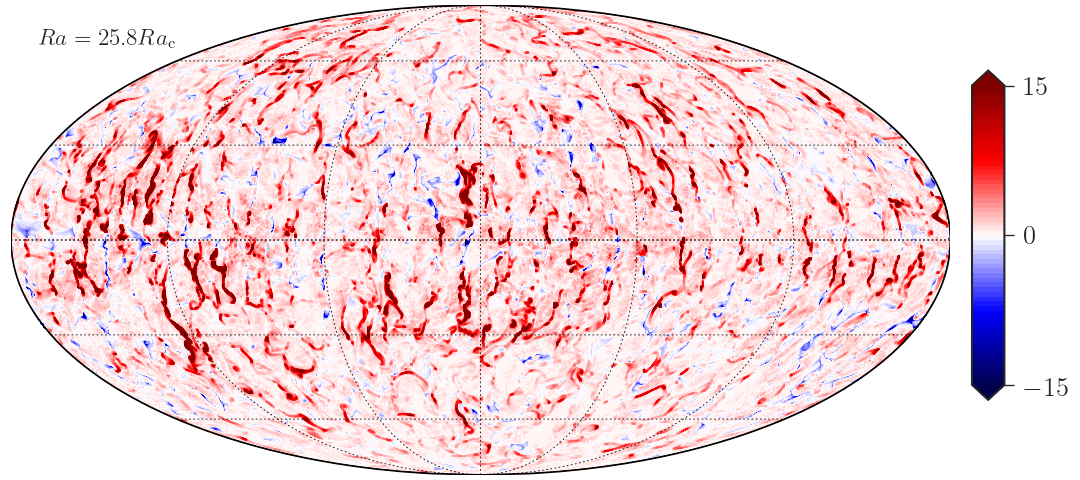}
     \caption{Mollweide projections of the surface radial magnetic field for a dipolar (top) and a multipolar (bottom) case with $\Nrho = 3$, corresponding to the run IDs FC22 and FC23, respectively. Red shades correspond to radial fields point outward and blue shades inward.
     }
     \label{fig:surfBr2}
\end{figure}

\begin{figure*}
     \centering
     \includegraphics[width=1.8\columnwidth]{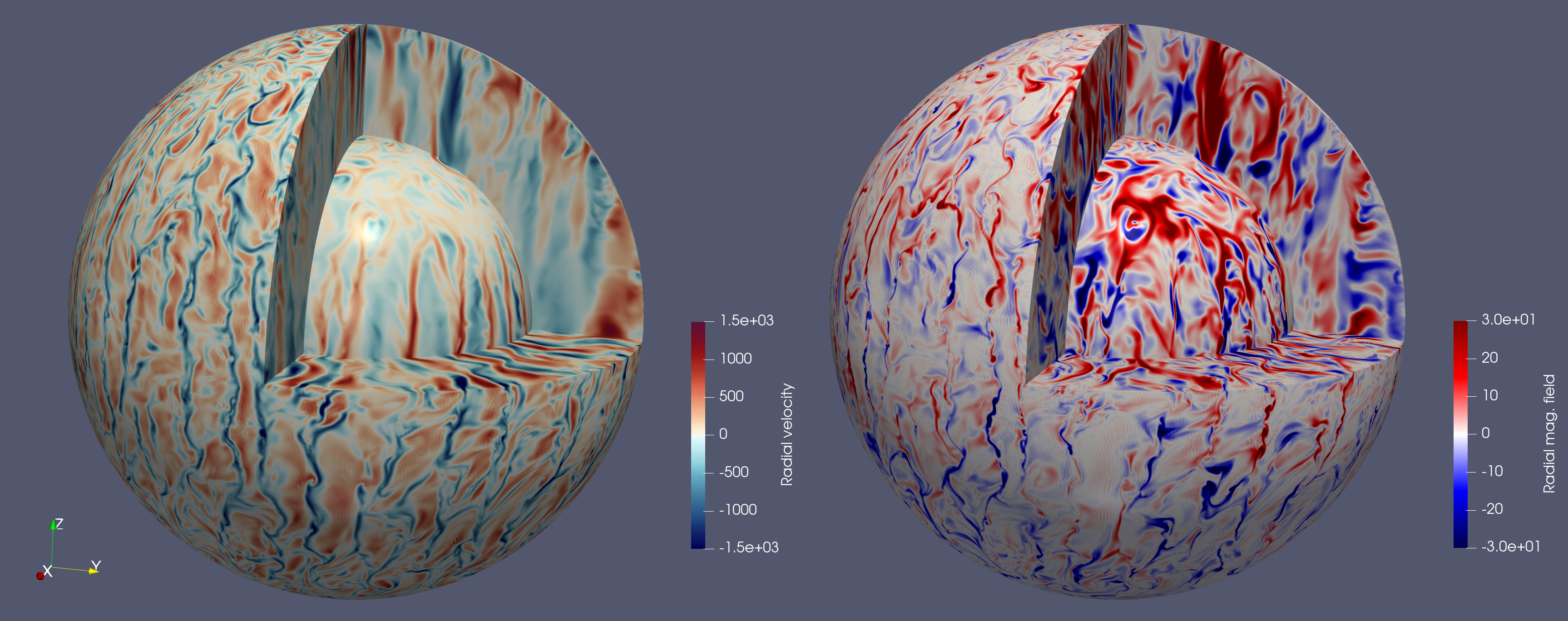}
     \caption{Snapshot of the radial velocity (left) and radial magnetic field (right) in the dipolar run shown in Figure\ref{fig:surfBr2} (FC22).}
     \label{fig:surfBr3}
\end{figure*}

The most striking result to emerge from the data is seen for the density contrast $N_\rho = 3$ (Figure~\ref{fig:fdipRa} c). Contrary to the other setups considered in this work, a multipolar dynamo is found close to the dynamo onset ($\ra = 1.3 \rac$). The dipolarity then shows a marked rise going from almost $0$ to $0.62$ as the forcing reaches about two times the critical Rayleigh number. Dipolar dynamos are then consistently sustained for a wide range of supercriticality until the morphology finally transitions to a multipolar configuration at $\ra\sim 25\rac$. Compared to the previous simulations of \citet{GDW12,GMD13} with $\prmag = 1$ and covering a parameter space of ${\ra < 5 \rac}$, we note that dipolar dynamos are kept for a much wider range of forcing. 
Comparing $\Nrho=1$ and $\Nrho=1.5$ simulations, we see that the range of $Ra$ numbers where the dipolar branch can be obtained shrinks as the density contrast increases. Although this result seems to reflect those of \citet{GDW12,GMD13} and \citet{J14}, who pointed out that dipolar dynamos would ultimately disappear for $\Nrho \gtrsim 2$, the strong dipoles obtained for $\Nrho = 3$ do not support this early conclusion. In fact, these results substantiate the previously unique simulation of \citet{YCM15}, which yielded a strong dipole ($\fdip \approx 0.55$) despite the high density contrast of $\Nrho = 5$. As argued by \citet{PR19}, one possibility is that the dipolarity loss found in previous works resulted from the restricted parameter space explored rather than being caused by a real modification of the dynamo mechanisms taking place in stars with different density contrasts. Indeed as we shall explore in Sec.~\ref{sec:fb}, our setup with $\prmag = 5$ increases the contribution of the Lorentz force to the force balance, sustaining dipolar dynamos even for stratification as high as $\Nrho = 3$. If anything, our simulations reinforce the idea that the regime of stability of dipolar dynamos depends on the parameter space explored \citep{RPD15} and provides evidence that sometimes higher stratification helps to sustain dipolar fields.

Figure~\ref{fig:surfBr2} shows the surface radial magnetic field for the last dipole before the transition (FC22) and the multipolar case after the collapse (FC23). Compared to the runs with $N_\rho=1$ (not shown here), smaller scales dominate the structure of the surface radial magnetic field in both cases. Indeed, a well-known effect of increasing the density stratification is to decrease the typical flow length scale, which in turn decreases the typical size of magnetic structures. We come back to this point when we discuss the scale at which the kinetic energy peaks in our simulations (see Sec.~\ref{sec:fb}). It is rather clear from this figure that a large-scale dipolar structure is present in the upper panel, with a positive North pole and negative South pole. On the contrary, in the bottom panel, the magnetic field is dominated by a salt and pepper like structure with the strongest field concentrations located in narrow bands more or less extended in latitude. Figure~\ref{fig:surfBr3} enables us to proceed to a closer inspection of the relationship between the flow and field morphologies. This figure shows a 3D rendering of the radial velocity field (left panel) and of the radial magnetic field (right panel) in the dipolar run shown in the top panel of Figure~\ref{fig:surfBr2}. It is rather clear from these 3D snapshots that narrow downwelling flows create intense magnetic flux concentrations, while broad upwelling flows diffuse the magnetic field. We also note that in this strongly stratified case and at this level of supercriticality $(\ra = 20.9 \, \rac)$, the amplitude of the convective velocities is strongest at the outer shell, as expected for strongly stratified systems. 
\subsection{The dipolar-multipolar transition} \label{sec:diptran}

Many studies interpreted the transition from the dipolar dynamos to multipolar dynamos in terms of the balance between inertia and Coriolis forces in the Navier-Stokes equation (Eq.~\ref{eq:mom_conserv}). A proxy to estimate this force ratio is the local Rossby number $\Ro$ introduced by \citet{CA06}. They suggested that the dipole-multipole transition is well captured by
\begin{equation} \label{eq:rossby}
        \Ro = \mean{\avg{\frac{u_\mathrm{rms}}{\Omega_o\dcz}\frac{\ell_u}{\pi}}}, \qq{where} 
    \ell_u = \frac{\sum_\ell \ell u_\ell^2}{\sum_\ell u_\ell^2}
\end{equation}
is the mean spherical harmonic degree of the flow. The global picture suggested that axial-dipole dominated solutions could only exist at low-Rossby numbers because of the ordering role played by the Coriolis force \citep[with typically $\Ro \lesssim 0.12$,][]{CA06}. Beyond this limit, the increased importance of inertia compared to Coriolis would cause the dipole collapse (with the star thus joining the multipolar branch). 

\begin{figure}
     \centering
     \includegraphics[width=\columnwidth]{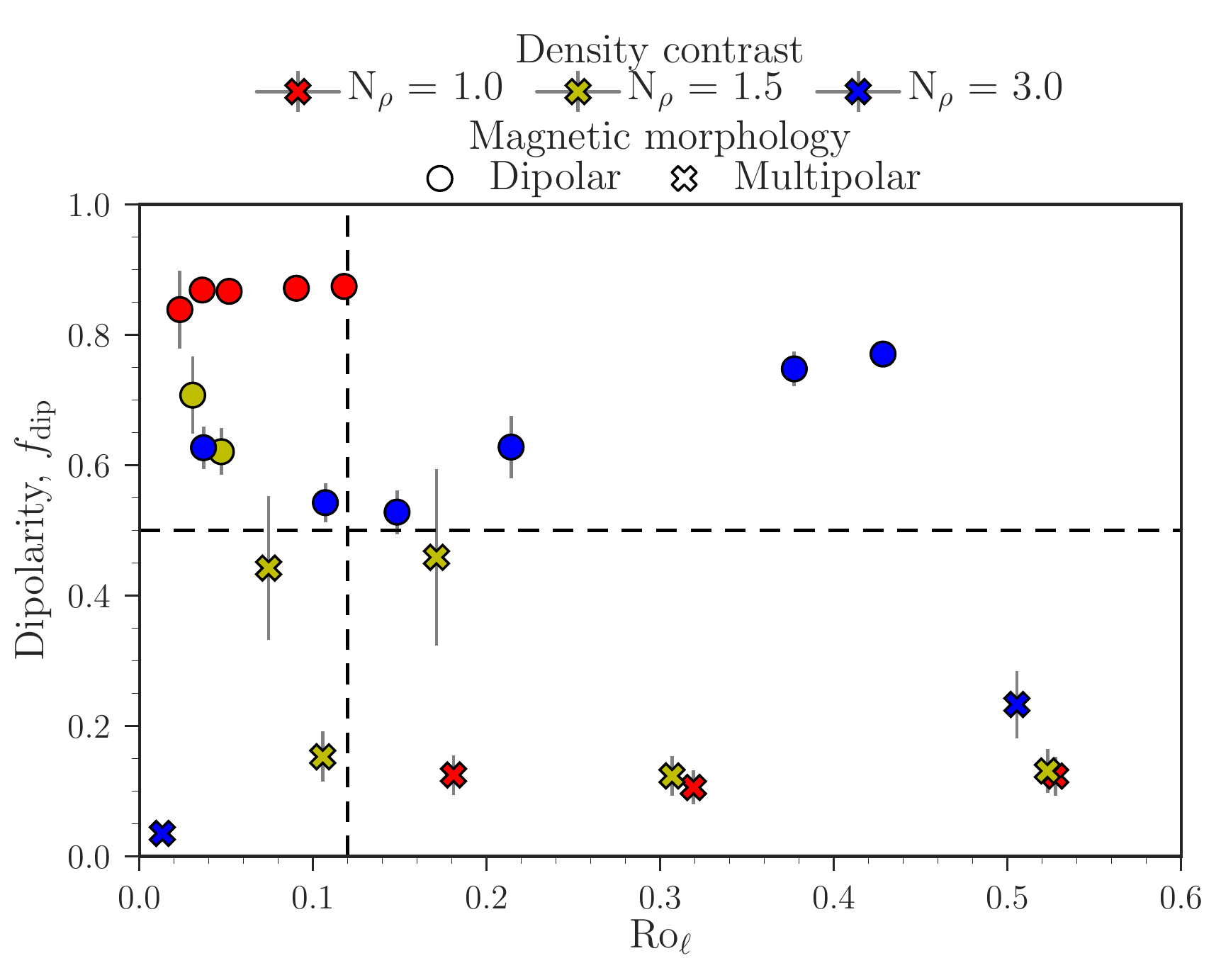}
     \caption{Surface dipolar fraction as a function of the local Rossby number $\Ro$ (Eq.~\ref{eq:rossby}). Colours group different levels of stratification (see legend), whereas symbols distinguish dipolar dynamos (circle) from multipolar dynamos (cross). The horizontal dashed black line marks the dipolar-multipolar transition, and the vertical one indicates the standard dipolar collapse predicted from geodynamo simulations \citep{CA06}.
     }
     \label{fig:fdip}
\end{figure}

We plot $\fdip$ as a function of $\Ro$ in Figure~\ref{fig:fdip}. Simulations with $\Nrho = 1$ display a dipolar-multipolar transition at $\Ro \sim 0.12$ (vertical dashed line), in agreement with Boussinesq results and arguments of \citet{CA06}. However, if we now turn to the runs with $\Nrho = 1.5$ or $3$, there is no clear evidence that $\Ro$ influences the dipole collapse. For these density contrasts, multipolar solutions are identified in the Rossby regime where mainly dipolar fields are predicted and vice-versa.

Perhaps one of the most interesting aspect evidenced by our simulations is that axial-dipole dominated simulations might display similar values of $\fdip$ regardless of whether it falls in the dipolar or multipolar branch as initially advised from Boussinesq simulations \citep{CA06}. Another key aspect is that dipolar solutions persist for large Rossby numbers precisely for the setup of highest density contrast ($\Nrho = 3$), which corresponds to the most realistic model in the stellar context. 

In an attempt to create a more general description for the dipolar transition, other proxies besides the Rossby number were explored in the literature to explain the possible causes for the dipole breakdown. As we discuss in Appendix~\ref{appendix:flow}, the change on the flow structure \citep{SKA12,GOD17} is not enough to explain the transition from dipoles to multipoles in our  numerical simulations. In particular, it seems that the magnetic morphology can only be described by a change on the flow arrangement when considering systems where the magnetic feedback on the flow is small/nonexistent (essentially behaving as a hydrodynamic flow).

Recently, Boussinesq simulations have shown that for systems in which the magnetic feedback is significant the relative importance of the Lorentz force in the Navier Stokes equation can describe the dipole breakdown \citep{MPG20,TGF21}. However, it is not clear whether those analyses still hold in anelastic dynamos. We explore next whether the balance between the forces entering the Navier-Stokes equation control the magnetic morphology in stratified systems.

\subsection{Force balance: inertia \textit{vs} Lorentz force} \label{sec:fb}
Following previous studies \citep{AGF17,SGA19,TGF21,GW21}, we compute the time-averaged root-mean-square (RMS) force spectra of the individual forces identified below 
\begin{equation*} \label{eq:mom_conserv2}
  \begin{split}
    &\underbrace{\ek\left[\pdv{\va{u}}{t}+(\va{u}\cdot\grad)\va{u}\right]}_{\text{Inertia}} +\underbrace{2\vu{e}_z\cross\va{u}}_{\text{Coriolis}} = \\ &-\underbrace{\grad(\frac{p}{\tilde{\rho}})}_{\text{Pressure}} + \underbrace{\frac{\ra\ek}{\pr} gs'\vu{e}_\mathrm{r}}_{\text{Buoyancy}} + \underbrace{\frac{1}{\prmag \tilde{\rho}}(\curl \va{B} )\cross\va{B}}_{\text{Lorentz}} + \underbrace{\frac{\ek}{\tilde{\rho}}\div S}_{\text{Viscous}} \qq*{.}
  \end{split} 
\end{equation*}
Here, time-averaged RMS force spectra are given by 
\begin{equation} \label{eq:forcelpeak}
     \mathcal{F}_\mathrm{RMS}(\ell) = \mean{\sqrt{\expval{\sum_{m=-\ell}^{\ell} \abs{\vec{F}_{\ell,m}(r,\theta,\phi,t)}^2}}}.
\end{equation}
where $\vec{F}_{\ell,m}$ is the vector spherical harmonic transform of the force at stake.

\begin{figure}
     \centering
     \includegraphics[width=\columnwidth]{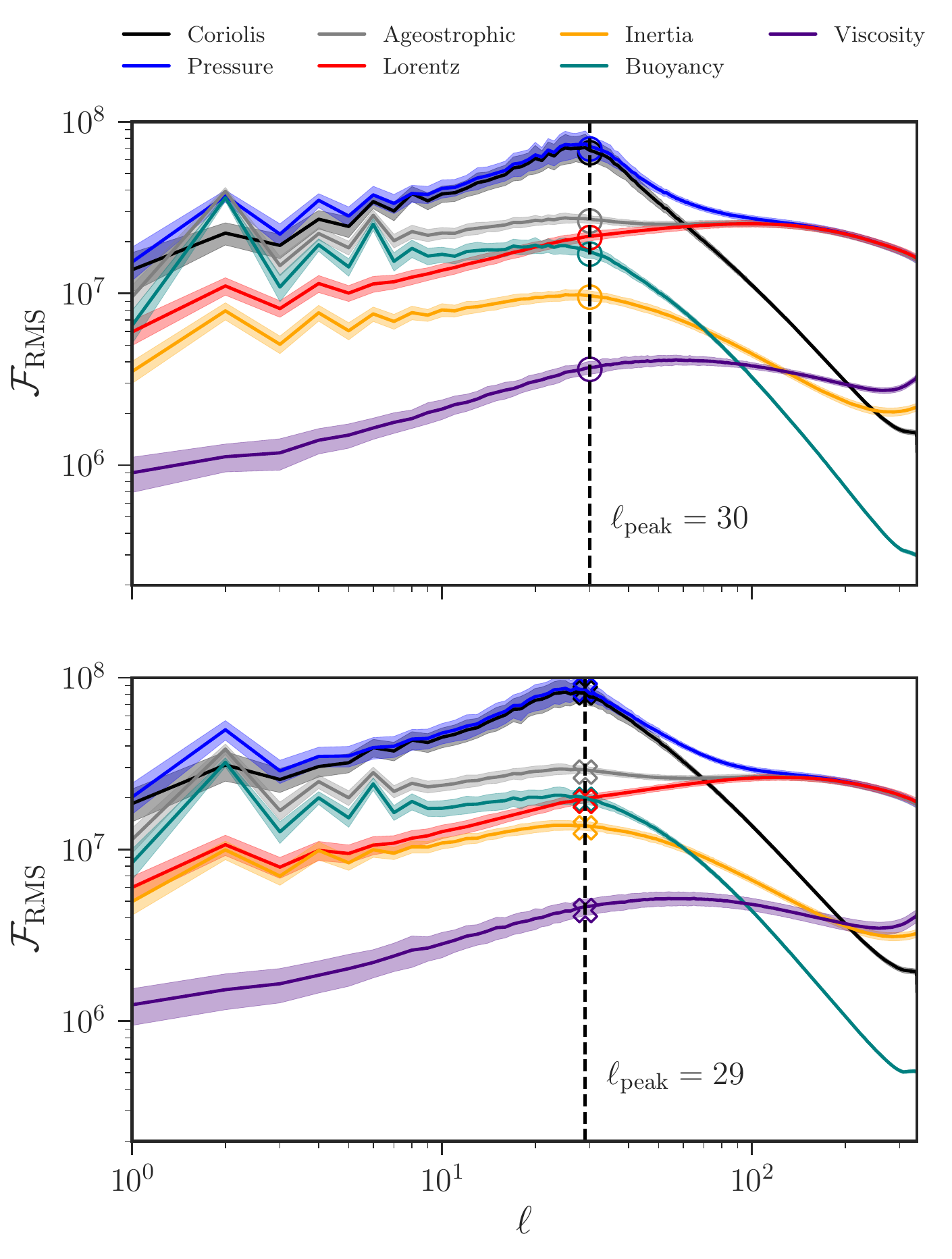}
     \caption{Force balance spectra for the same dipolar (top) and multipolar (bottom) models shown in Figure~\ref{fig:surfBr2}. Solid lines correspond to time-averaged force spectra, with colours representing the different forces entering the Navier-Stokes equation. Shaded regions represent one standard deviation from the time-averaged value. The vertical dashed line marks the integral scale $\ell_\mathrm{peak}$ defined in Appendix~\ref{appendixB}.}
     \label{fig:FBspec}
\end{figure}

Figure~\ref{fig:FBspec} illustrates the force balance spectra for a dipolar and a multipolar run with $\Nrho = 3$ (corresponding to the same runs shown in Figure~\ref{fig:surfBr2}). Both models display forces whose respective contributions vary depending on the spatial scale. At scales up to $\ell \sim 40$, the Coriolis (black) and pressure (blue) forces balance each other at first order resulting in a quasi-geostrophic balance \citep[QG, for further details, see][]{C18}, whereas buoyancy (green), Lorentz (red), and inertial (yellow) forces show a marginal contribution at second-order. On the other hand, at small scales ($\ell \gtrsim 40$) the Lorentz force becomes dominant and starts to balance the pressure force in the place of the Coriolis force. Comparing both models, we can identify an increase in the inertial contribution from the dipolar to the multipolar case, with the inertial force reaching values comparable to the Lorentz force in the latter.

To track the relative contribution of each force in our parametric study, we look for a particular length scale $\ell_\text{peak}$ defined as the dominant scale of the convective flow \citep[for more details on its calculation, see appendix~\ref{appendixB} and][]{SGA21}. The values of $\ell_\text{peak}$ are given in Table~\ref{tab:runs} for each simulation. We note here that the impact of the density stratification is reflected in the strong increase of $\ell_\text{peak}$ with $N_\rho$. Indeed, from $N_\rho=1$ to $N_\rho=3$, $\ell_\text{peak}$ is typically multiplied by a factor 2. We now compute the RMS forces at the integral scale $\ell_\text{peak}$, namely, Coriolis force $\Fcor$, pressure gradient force $\Fpre$, buoyancy (or Archimedes) force $\Fbuo$, Lorentz force $\Flor$, inertial force $\Finer$, and the viscous force $\Fvis$.

\begin{figure}
     \centering
     \includegraphics[width=\columnwidth]{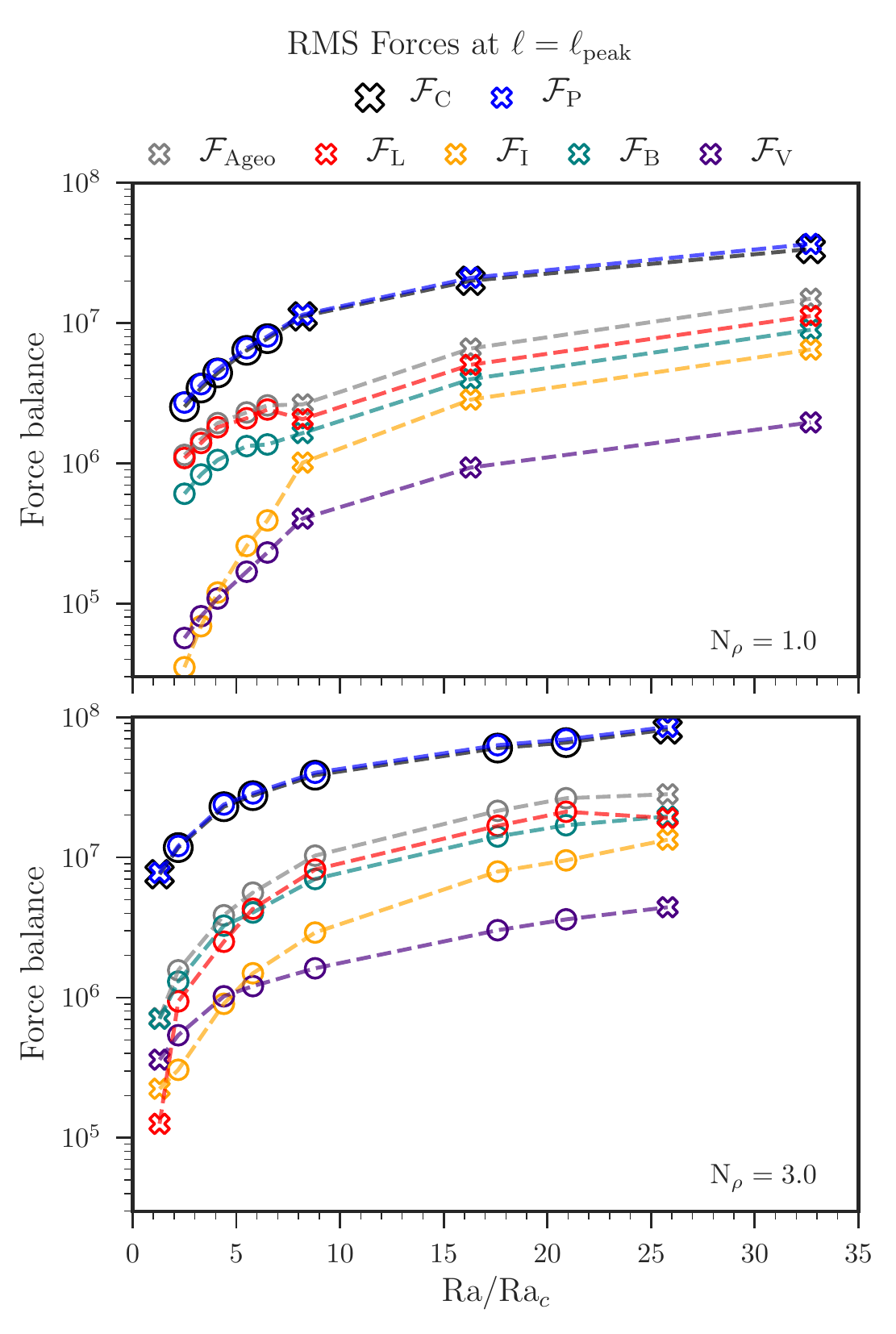}
     \caption{Force contributions at the integral scale $\ell_\text{peak}$ (Eq.~\ref{eq:forcelpeak}) as a function of $\ra/\rac$. Top and bottom panels show runs with N$_\rho = 1$ and $3$, respectively}.
     \label{fig:FBra}
\end{figure}

Figure~\ref{fig:FBra} shows these forces as a function of $\ra/\rac$ for models with $\Nrho = 1$ and 3. While the entire data set features a QG balance at first order, the ageostrophic part of the Coriolis force, defined as $\mathcal{F}_\mathrm{Ageo} = \abs{\Fcor - \Fpre}$, enters a second-order force balance that varies depending on $\Nrho$ and $\ra$.

For $\Nrho = 1$ (top panel), we identify two kinds of second-order balance depending on the Rayleigh number. At $\ra < 7 \rac$, the ageostrophic Coriolis force is balanced by $\Flor$ and $\Fbuo$ forces, which dominate over $\Finer$ and $\Fvis$ by roughly an order of magnitude. This flow state, devised by \citet{D13}, is frequently referred to as the quasi-geostrophic Magneto-Archimedean-Coriolis (QG-MAC) balance, and it has been obtained in geodynamo models \citep{YGC16,AGF17,SJN17} and in anelastic models of gas giant planets \citep{GW21}. At $\ra > 7 \rac$, inertial forces become important and contribute to the second-order balance of the Navier-Stokes equation. We observe that the breakdown of the dipole occurs at this point. The role played by inertia in destabilizing dipoles was likewise found before in Boussinesq simulations \citep[e.g.,][]{SJ06,CA06}.    

Similar conclusions can be drawn for the $\Nrho = 3$ data set (bottom panel), with the main difference relying on the isolated multipolar solution at $\ra = 1.3 \rac$, i.e., very close to the convective onset. Among the entire set of simulations performed, this case is the only one that does not display a dominant Lorentz contribution to the flow dynamics. Instead, it yields a strong contribution of $\Fbuo$ and a marginal one of $\Fvis$. This flow adjustment is often called quasi-geostrophic Viscous-Archimedean-Coriolis (QG-VAC) balance \citep{YGC16,SGA21}. The QG-VAC balance is quickly destroyed as turbulence builds-up due to a sharp rise in the $\Flor$ with $\ra$. One of the main conclusions we can extract from Figure~\ref{fig:FBra} is that with this stratification, dipolar dynamos prevail for much higher $\ra/\rac$ than for the less stratified cases. The transition in the surface field morphology is indeed seen at $\ra = 25.8 \rac$. Akin to what has been described for $\Nrho = 1$, the morphology transition occurs as the gap between $\Flor$ and $\Finer$ decreases. This finding suggests that, in the Lorentz force dominated regime, the effect of the density stratification is to increase the level of turbulence at which inertial forces become comparable to the Lorentz forces.

\begin{figure}
     \centering
     \includegraphics[width=\columnwidth]{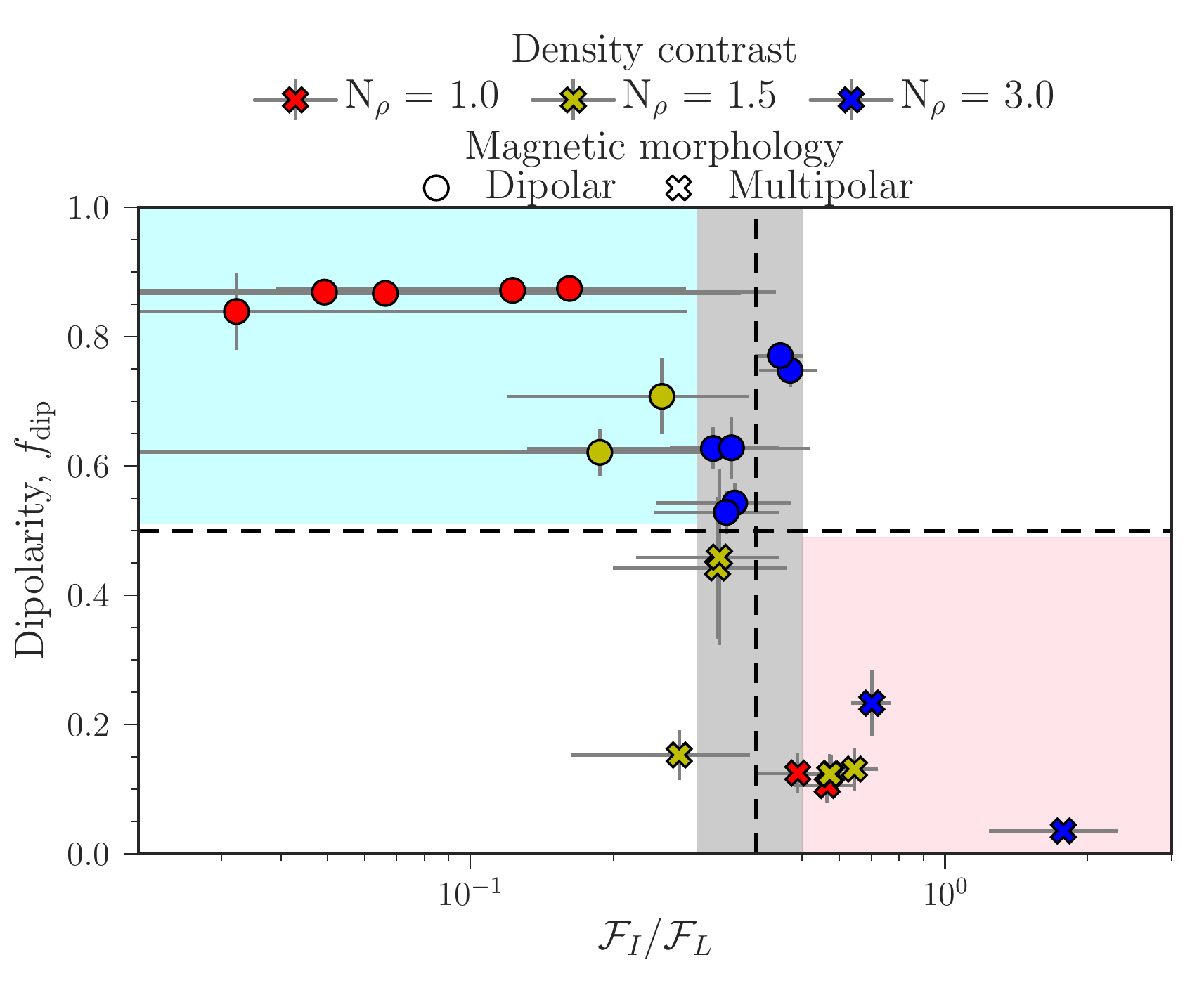}
     \caption{Surface dipolar fraction as a function of the ratio between inertia and Lorentz force at the integral scale. Symbols are defined as in Figure~\ref{fig:fdip}. The vertical dashed black line indicates the tentative threshold ${\Finer/\Flor = 0.4}$ for the dipole breakdown. The error bars correspond to one standard deviation about the time-averaged quantities. Shaded areas indicate the dipolar (cyan) and multipolar (coral) branches proposed in this work. 
     }
     \label{fig:FBdip}
\end{figure}

To test the hypothesis that the importance of inertia in the 2nd-order force balance is the main factor responsible for destabilising dipolar solutions, we plot in Figure~\ref{fig:FBdip} the dependence between $\fdip$ and $\Finer/\Flor$ for the three setups considered in this work. Dipolar and multipolar branches are identified using this proxy. We find that simulations with ${\Flor \gg \Finer}$ develop strong dipolar dynamos, while a sharp transition to multipolar dynamos is obtained as inertia increases in intensity. A tentative description for the dipolar-multipolar transition gives ${\Finer/\Flor \simeq 0.4}$ (vertical dashed line). It follows that $\Finer/\Flor$ provides a more unified view of the dipolar-multipolar transition than $\Ro$ (Figure~\ref{fig:fdip}), independently of the density contrast $N_\rho$. This result agrees with those of \citet{MPG20} and \citet{TGF21}, who also found that the competition between inertial and Lorentz forces can capture the dipole collapse in Boussinesq simulations. We thus confirm that these results still hold in stratified systems, and even argue that the transition may occur at larger levels of turbulence for strongly stratified cases, opening the possibility that stars harbouring strong dipoles may indeed operate in this Lorentz force-dominated regime. 

\subsection{Possible proxies for stellar observations}
\label{sec:criteriaObs}
\subsubsection{Energy distribution}\label{energy_proxy}

Following \cite{TGF21}, we now try to look for an alternative quantity to the ratio ${\Finer/\Flor}$ that is more accessible to observations and yet incorporates the physics behind the dipole collapse. To establish this new measure, we use the kinetic energy stored in the convective motions ($E_K$) as a proxy of the inertial force and the magnetic energy ($E_M$) as a proxy of the Lorentz force. The rough approximation of ${\Finer/\Flor}$ is then given by the time and volume-averaged energy ratio
\begin{equation} \label{eq:ekem}
    \frac{E_K}{E_M} = \ek\prmag \mean{\frac{\avg{\tilde{\rho}\va{u}^2}}{\avg{\va{B}^2}}}.
\end{equation}

\begin{figure}
     \centering
     \includegraphics[width=\columnwidth]{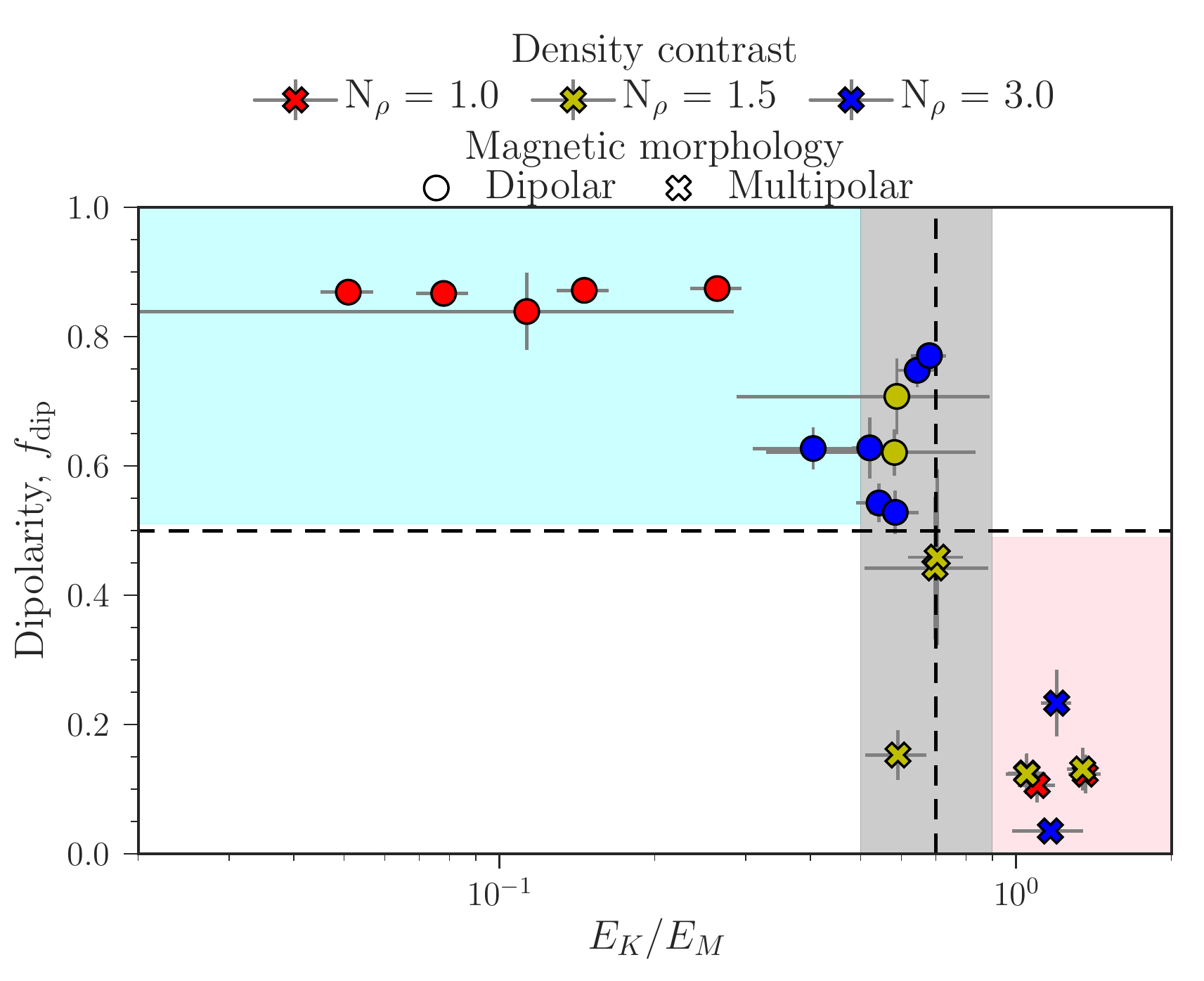}
     \caption{Surface dipolar fraction in terms of the ratio of the time and volume-integrated kinetic energy stored in the convective motions and magnetic energy. The vertical dashed black line indicates the tentative threshold ${E_K/E_M = 0.7}$ for the dipole collapse. Shaded areas indicate the dipolar (cyan) and multipolar (coral) branches proposed in this work. 
     }
     \label{fig:Energydip}
\end{figure}

Figure~\ref{fig:Energydip} shows the dipolarity in our simulations as a function of this new proxy $E_K/E_M$. We find dipolar morphologies at low-$E_K/E_M$ and complex multipolar morphologies below equipartition (i.e., $E_K/E_M > 1$). These findings suggest that the energy ratio can likewise capture the dipolar-multipolar transition. It stands out that the energy ratio $E_K/E_M$ in the dipolar cases with $N_\rho = 1$ are significantly smaller than those obtained for the other density contrasts. This behaviour reflects what was already seen in Figure~\ref{fig:FBdip} using the force ratio, providing further evidence that $\Finer/\Flor$ and $E_K/E_M$ are indeed correlated. This occurs because the magnetic energy generated in these models is 2-6 times larger than the ones reached by other dipolar simulations in the same range of supercriticality (and hence with similar $E_K$). The shaded areas in Figure~\ref{fig:Energydip} show the tentative dipolar (cyan) and multipolar (coral) branches, along with a transitional region (grey) set to match the uncertainties of $E_K/E_M$ in the runs falling in the transition. From the data, we derive that the dipole breakdown occurs around ${E_K/E_M \simeq 0.7}$ (vertical dashed line). 

As we will discuss in  Section~\ref{sec:conclusions}, one advantage of the energy ratio is that we can use stellar observations to estimate $E_K/E_M$ at the stellar surface. Such an observational quantity is not strictly speaking identical to the definition in Eq.~\ref{eq:ekem}, and we could instead compute $E_K/E_M$ at the surface of our numerical simulations. However, the surface of numerical simulations differs from the surface of stars because boundary conditions constrain the field and flow. Moreover, 3D anelastic dynamo simulations better reflect the physics of the stellar convective envelope when excluding the outer few per cent of the radial domain as the anelastic approximation loses its validity at the stellar surface. For those reasons, we believe that volume-averaged energies are more adequate when drawing a parallel between numerical simulations and observations in Section~\ref{sec:conclusions}.

\subsubsection{Differential rotation} \label{sec:dyngen}
Stellar observations can give access not only to the surface magnetic fields in stars but also on some flow characteristics, like the surface differential rotation \citep[e.g.,][]{DMP08,MDP08}. Since we can measure in detail the differential rotation obtained in our calculations, we propose here to determine the amplitude and sign of the latitudinal differential rotation obtained in our dipolar and multipolar dynamo simulations. This will be used mostly for a comparison to the observations discussed in the following section.

Although numerical studies usually compute the latitudinal shear as the difference between the angular velocity at the equator minus an arbitrary latitude close to the poles, this parameter strongly depends on the chosen polar latitude as fast zonal flow variations may exist. Therefore, we compute the relative surface shear using a less dependent definition based on the difference between the angular velocity averaged on the near-surface layer (NSL) at equatorial regions and polar regions:
\begin{equation}\label{eq:shear}
\chi_\Omega = \frac{\langle\mean{\Omega}\rangle_{\mathrm{NSL}, |\theta|<40^o} - \langle\mean{\Omega}\rangle_{\mathrm{NSL}, 40^o<|\theta|<80^o}}{\Omega_o}.
\end{equation}
Here, we define as NSL the outer shell with thickness 0.05 $\rout$ and we exclude high latitudes with $|\theta|>80^o$ from our computations (where small scale features are observed but should likely average out if considering longer time averages). 

\begin{figure}
     \centering
     \includegraphics[width=\columnwidth]{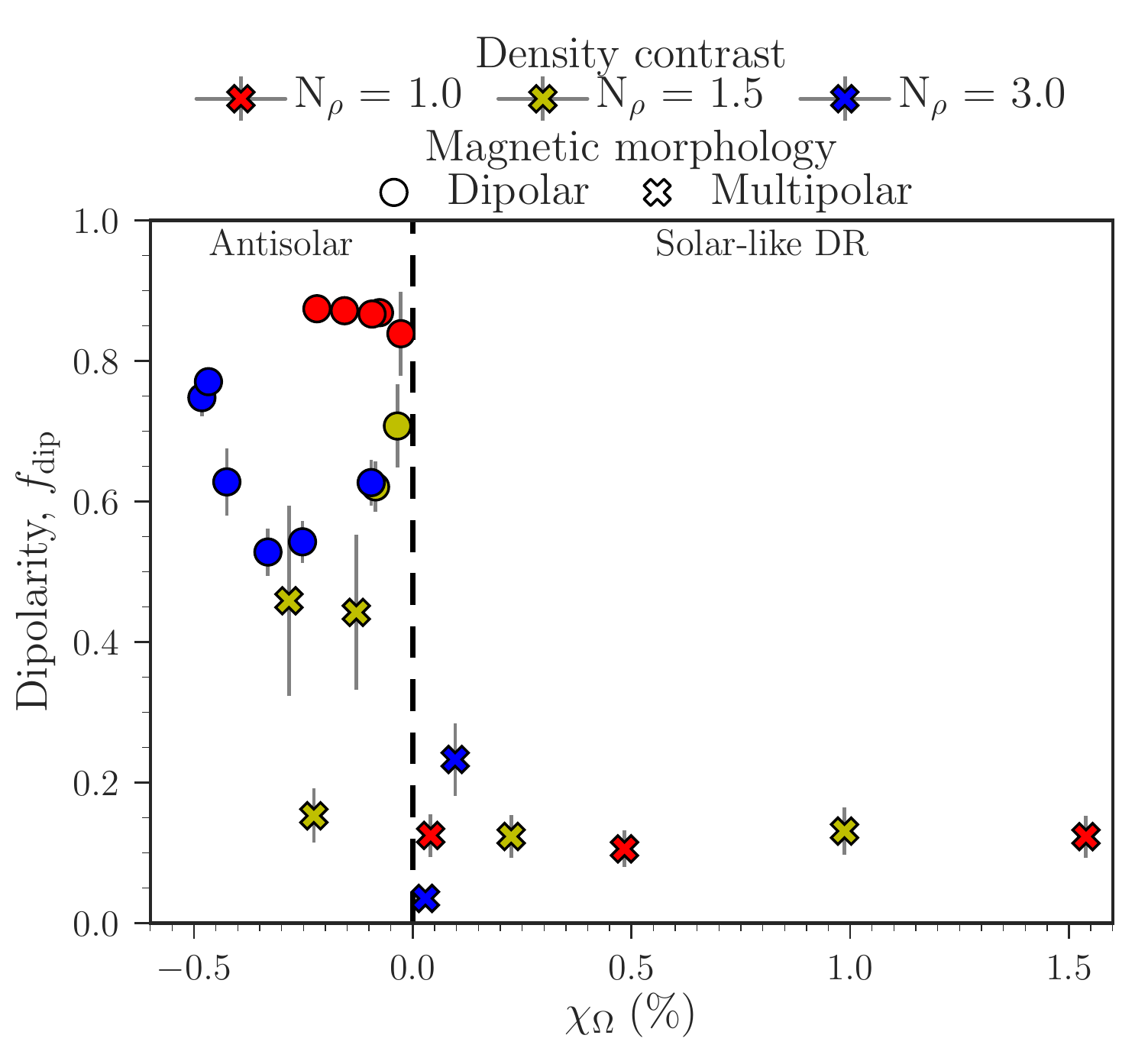}
     \caption{Dipolarity as a function of the differential rotation measured at the surface. The dashed vertical line represents a solid body rotation using our shear definition in Eq.~\ref{eq:shear}. Simulations with negative (positive) $\chi_\Omega$ display antisolar differential rotation profiles, while those with positive $\chi_\Omega$ have solar-like differential rotation profiles.}
     \label{fig:DRsurf}
\end{figure}

Figure~\ref{fig:DRsurf} shows the dipolarity as a function of the relative latitudinal shear at the near-surface layer (cf. Eq.~\ref{eq:shear}). We want to emphasize that a non-dimensional quantity is used here to quantify the shear since the value of $\Omega_0$ in physical units is not set \emph{a priori} in our simulations. However, note that the same Ekman number (and thus the same $\Omega_0$) is used in all simulations so that the trend would be similar if only the numerator of Eq.~\ref{eq:shear} was used as the x-axis of Figure~\ref{fig:DRsurf}. The first striking feature is that all simulations exhibit a rather weak level of  differential rotation with $\chi_\Omega < 2\%$. This quenching on the differential rotation can be understood because magnetic stresses are always active in our calculations as Lorentz forces significantly impact the flow \citep{COG99,B02}. Another important result is that the level of surface differential rotation is not negligible in dipolar cases, especially at $N_\rho=3$, compared to the multipolar ones. However, an important difference between dipolar and multipolar simulations is the differential rotation sign. Figure \ref{fig:DRsurf} indeed reveals that all simulations with dipole dominated morphology build an antisolar differential rotation profile. We find that non-negligible relative shears exist in our dipolar cases, with $\chi_\Omega$ ranging from $-0.57$ to $-0.03\%$. We note that these antisolar profiles were also observed in the geodynamo simulations of \citet{A05} and only illustrate the fact that the Lorentz force plays a significant role here in the angular momentum transport. On the other hand, solar-like differential rotation profiles only show up in the multipolar simulations. The only three multipolar cases developing antisolar profiles are those with $N_\rho = 1.5$, whose dipoles are either reversing or excursioning. The equatorial acceleration seen in the multipolar cases is consistent with the fact that it is only in this situation that inertia becomes comparable to Lorentz forces, as discussed in Section~\ref{sec:diptran}. This finding is in line with the non-magnetic simulations of \cite{GYM14}, where solar-like profiles are found when Reynolds stresses are enhanced. Indeed, the Reynolds stresses, associated with inertial forces, are known to be responsible for the equatorial acceleration of the flow \citep{M05}. They thus need to be significant enough to counteract the angular momentum transport by magnetic fields. When considering the multipolar simulations with solar-like differential rotation, we find that equatorial regions indeed accelerate, with values going up to $1.5\%$.

\section{Discussion and Conclusions} \label{sec:conclusions}
This paper explored through 3D dynamo simulations the physical mechanisms responsible for controlling the magnetic morphology of large-scale fields in partly convective cool stars. To address this point, we carried out 23 simulations of a spherical convective rotating shell with a radius ratio of 0.6 between the bottom and the top of the shell. Our modelling strategy follows recent geodynamo studies of \citet{MPG20} and \citet{TGF21}, who suggested that having a significant Lorentz force contribution in the force balance when simulating convective dynamos could modify conclusions about the magnetic morphology. However, unlike their study, we considered a fluid layer with a density contrast between the top and bottom of the convective zone to model conditions applicable to stellar interiors. 

Our simulations demonstrate for the first time that axial dipole dominated solutions can be achieved at large Rossby numbers in stratified systems (up to $\Ro=0.4$). Even more important maybe is the fact that these dipoles at high $\Ro$ are obtained for simulations with a large density contrast between the top and bottom of the convective zone, at $N_\rho=3$. This finding differs from previous numerical studies suggesting that dipolar dynamos would only exist at low-Rossby numbers \citep[e.g.,][]{CA06,GDW12} and that strong stratification may make it more difficult for dipoles to survive.
In the same vein, \citet{RPD15} have also suggested that dynamos may be obtained for strong stratification, but we here extend the validity of their result to $Ro > 0.1$. In particular, it represents an important step towards the understanding of the magnetic morphology of stars, as strong axial dipoles have been likewise observed in some stars with $\Ro > 0.1$, e.g., \mbox{TYC~5164-567-1} \citep[$\fdip=0.77$; ][]{FPB16}, V439~And \citep[$\fdip=0.60$; ][]{FPB16}, HD~6569 \citep[$\fdip=0.53$;][]{FBP18}, and CE~Boo \citep[$\fdip=0.76$;][]{DMP08}. We note that we also find solutions at $N_\rho=1.5$ with flipping or excursioning dipoles, producing measures of the dipolar fraction which can significantly vary in time. This could potentially be reminiscent to the strong variations in the dipolar and quadrupolar modes observed in the Sun \citep{DBH12} or other solar-like stars over their magnetic cycle \citep[e.g.,][]{PDS08,BLJ18}, all falling under the high Rossby regime.
 
Taken together, our parameter survey evidenced that the Rossby number cannot capture the transition in the surface field morphology when the Lorentz force is strong. We explored the possible mechanisms causing the axial dipole collapse using the relative amplitude of the axial dipole at the surface to measure the magnetic morphology in our simulations (cf. Eq.~\ref{eq:fdip}). From the investigation of the flow configuration, there was no evidence of its influence on the magnetic morphology. These findings can be understood by the significant back reaction of the magnetic field on the flow through the Lorentz force. As argued in the early study of \citet{GOD17}, the flow configuration only emerges as a good proxy of the magnetic morphology when the flow transitions are similar to those observed in hydrodynamical simulations. Indeed the force balance analysis shows a significant Lorentz force contribution to the flow dynamics in our calculations.

An important finding that emerged from the force balance study is that the ratio between the inertial and magnetic forces can describe the dipole-multipole transition of dynamo models with a background density contrast. We found that the dipole branch is recovered when the Lorentz force dominates over the initial force, with the transition to multipolar branch occurring around $\Finer/\Flor \simeq 0.4$. Similar to the conclusions obtained in past anelastic studies, it remains valid that the increased influence of inertia on the flow is responsible for destabilizing the axial dipoles. However, our work shows that instead of the traditional comparison with the Coriolis force (through the Rossby number), it is the relative importance of inertia compared to the Lorentz force that controls the transition if the magnetic back reaction on the flow is strong. With similar conclusions drawn by recent geodynamo simulations with $N_\rho= 0$ \citep{MPG20}, $\Finer/\Flor$ seems to emerge as a reliable predictor of the magnetic field morphology of stars and planets. 

Because a direct estimate of the actual forces at play is not practical in stellar interiors, we explored an alternative proxy based on the ratio of kinetic to magnetic energies \citep{TGF21}. The investigation of $E_K/E_M$ revealed dipolar and multipolar branches confirming the ability of $E_K/E_M$ to describe the dipole collapse in numerical simulations \citep[early proposed by Boussinesq simulations;][]{KC02,TGF21}. From our data set, we found that stratified systems emerge as multipolar dynamos whenever $E_K/E_M \gtrsim 0.7$. 

To tentatively test this proxy with observations, we gathered from the literature partly-convective stars with large-scale surface magnetic fields reconstructed using the Zeeman-Doppler imaging technique \citep[for details of the technique see, e.g.,][]{DSC97,DB97,DHJ06}. Given that our simulations correspond to a convective shell spanning the outer $40\%$ of the radial domain, we focused on partly convective M dwarfs with masses ranging from 0.38 to 0.60\,$M_\odot$, whose convective zones are expected to feature radius ratios (between the bottom and top of the convective zone) ranging from 0.50 to 0.66 \citep[estimated with the ATON code, described in][]{LVD06}, i.e., with roughly the same extension as those modeled in our simulations. We consider for consistency the homogeneous sample of stars published by \citet{DMP08} and \citet{MDP08}, which had their surface magnetic maps reconstructed with the same Zeeman-Doppler imaging code. We find eight stars obeying the mass condition described above: GJ~182, DT~Vir, DS~Leo, GJ~49, OT~Ser, CE~Boo, AD~Leo, and EQ~Peg~A. We also take into account multiple magnetic field reconstructions existent for DT~Vir, DS~Leo, and OT~Ser (with each star being observed at two different epochs). 

From their magnetic surface maps, we directly derive $E_M$ based on the averaged surface magnetic field ($B_\mathrm{rms}$) and a modified dipolarity that is comparable to our definition in Eq.~\ref{eq:fdip} but with a maximum spherical harmonic degree that varies depending on the spatial resolution achieved for each star (typically $\ell_\mathrm{max}$ ranged from 6 to 10). We find that under our morphology classification CE Boo, AD Leo, and EQ~Peg~A fall in the criteria of dipolar dynamos ($\fdip$ = 0.76, 0.57, and 0.57, respectively), while the other stars harbour a multipolar dynamo. Because observations only have access to the magnetic energy at the surface, we accordingly estimate the surface kinetic energy $E_K$ to compute the energy ratio of each star. We use published values of mass $M_\star$ and radius $R_\star$ present in the original Zeeman-Doppler imaging study. We adopt a rough approximation for the turbulent velocity $u_\mathrm{rms} = R_\star/\tau_c$ and photospheric density $\rho_{\star,\mathrm{pho}} = \frac{\bar{\rho}_\star}{\bar{\rho}_\odot}\rho_{\odot, \mathrm{pho}}$, where $\tau_c$ is the convective turnover time derived with the empirical relationships based in the stellar mass $M_\star$ \citep{WNW18},
$\bar{\rho}_{\odot, \star} = M_{\odot, \star}/(4\pi R_{\odot, \star}^3/3)$ is the mean density, and $\rho_{\odot,\mathrm{pho}}\approx 10^{-6}\,g\,cm^{-3}$ is the Sun's photospheric density \citep{BS05}. We thus estimate
\begin{equation} \label{eq:empiralEkEm}
    \frac{E_K}{E_M} = \frac{\rho_{\star,\mathrm{pho}} u_\mathrm{rms}^2}{2} \frac{8\pi}{B_\mathrm{rms}^2} \approx \frac{4\pi}{B_\mathrm{rms}^2} \left(\frac{M_\star}{M_\odot} \right) \left(\frac{R_\odot}{R_\star}\right)^3 \left(\frac{R_\star}{\tau_c} \right)^2 \rho_{\odot,\mathrm{pho}}.
\end{equation}

\begin{figure}
     \centering
     \includegraphics[width=\columnwidth]{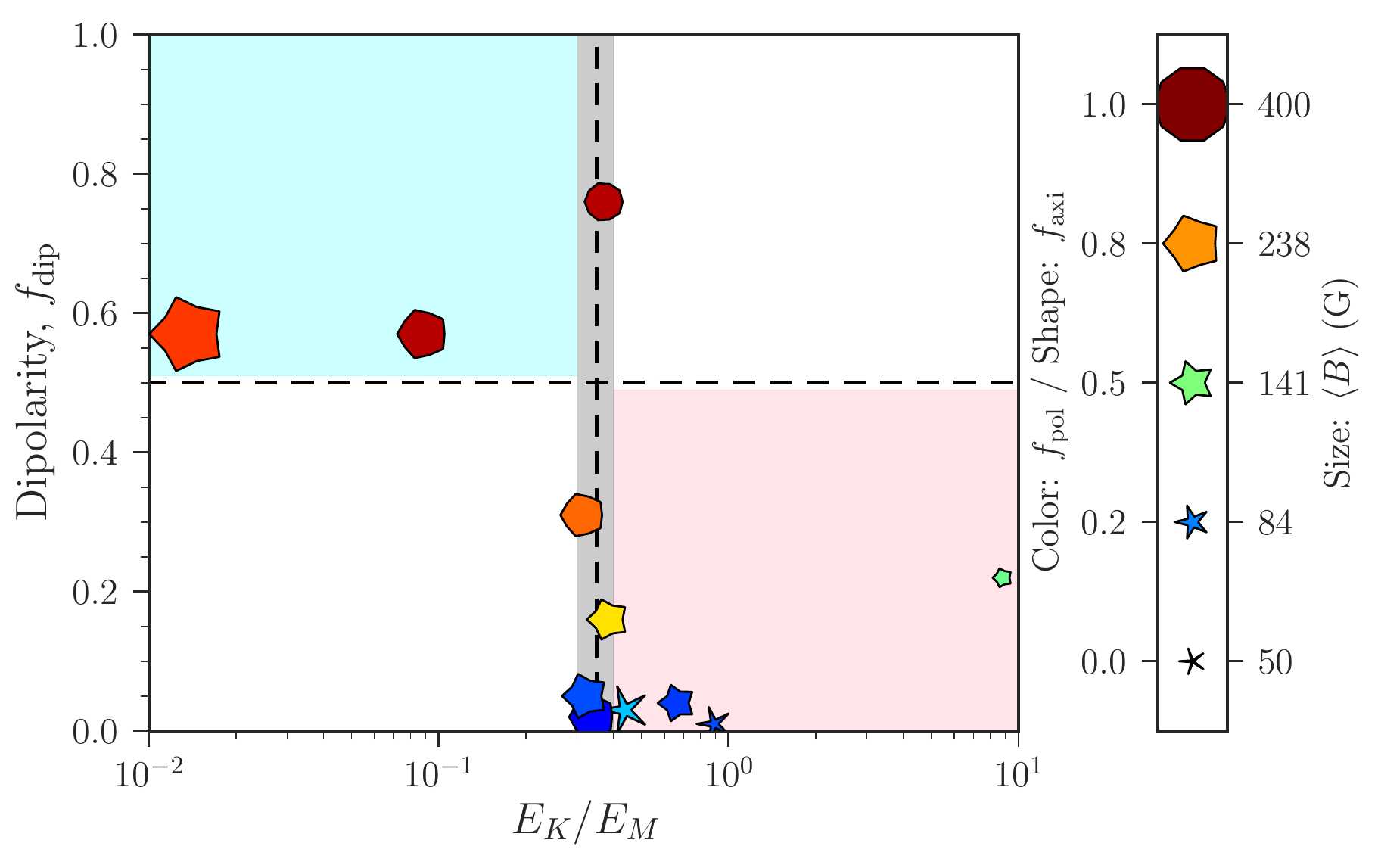}
     \caption{Observational counterpart of Figure~\ref{fig:Energydip}. Symbols show the magnetic properties of the M dwarfs derived with the Zeeman-Doppler imaging technique  \citep{DMP08,MDP08}. The symbol size correspond to the field strength at the surface $\langle B \rangle$, the shape corresponds to the degree of axisymmetry of the magnetic field, and colors represent the amount of energy stored in the poloidal field. Shaded areas are similar to Figure~\ref{fig:Energydip}, with cyan representing strong dipoles axisymmetric fields (top left) and coral the multipolar non-axisymmetric fields (bottom right). However, we use a dipole-multipole transition of $E_K/E_M = 0.35$ (vertical dashed line) that is lower than the one obtained with simulations ($E_K/E_M = 0.7$). }
     \label{fig:dipOBS}
\end{figure}

Figure~\ref{fig:dipOBS} illustrates the magnetic properties of M~dwarfs as a function of the energy ratio computed with Eq.~\ref{eq:empiralEkEm}. The sharp transition in the magnetic morphology is apparent from this plot. We find that M dwarfs with $E_K/E_M \lesssim 0.35$ have surface large-scale magnetic fields that are mostly poloidal and with strong axisymmetric dipoles. In contrast, M dwarf stars with higher energy ratios $E_K/E_M$ host large-scale fields with strong toroidal fields and weak axial dipoles. We infer a dipolar-multipolar transition around $E_K/E_M \simeq 0.35$ (dashed vertical line) from the observational data. As we considered volume-averaged energies instead of surface-averaged energies in our simulations (see details in Section~\ref{energy_proxy}), it is not surprising that observations show a dipole collapse at a different value than the one predicted from our simulations. Despite that, it is encouraging to see that an energy ratio proxy also seems to describe the transition in the magnetic morphology of M~dwarfs. Future simulations with different sizes of the convective envelope will help assess whether the dipole collapse is sensitive to this parameter and, therefore, if it is a potential source of uncertainties when determining the $E_K/E_M$ threshold.

\begin{figure}
     \centering
     \includegraphics[width=\columnwidth]{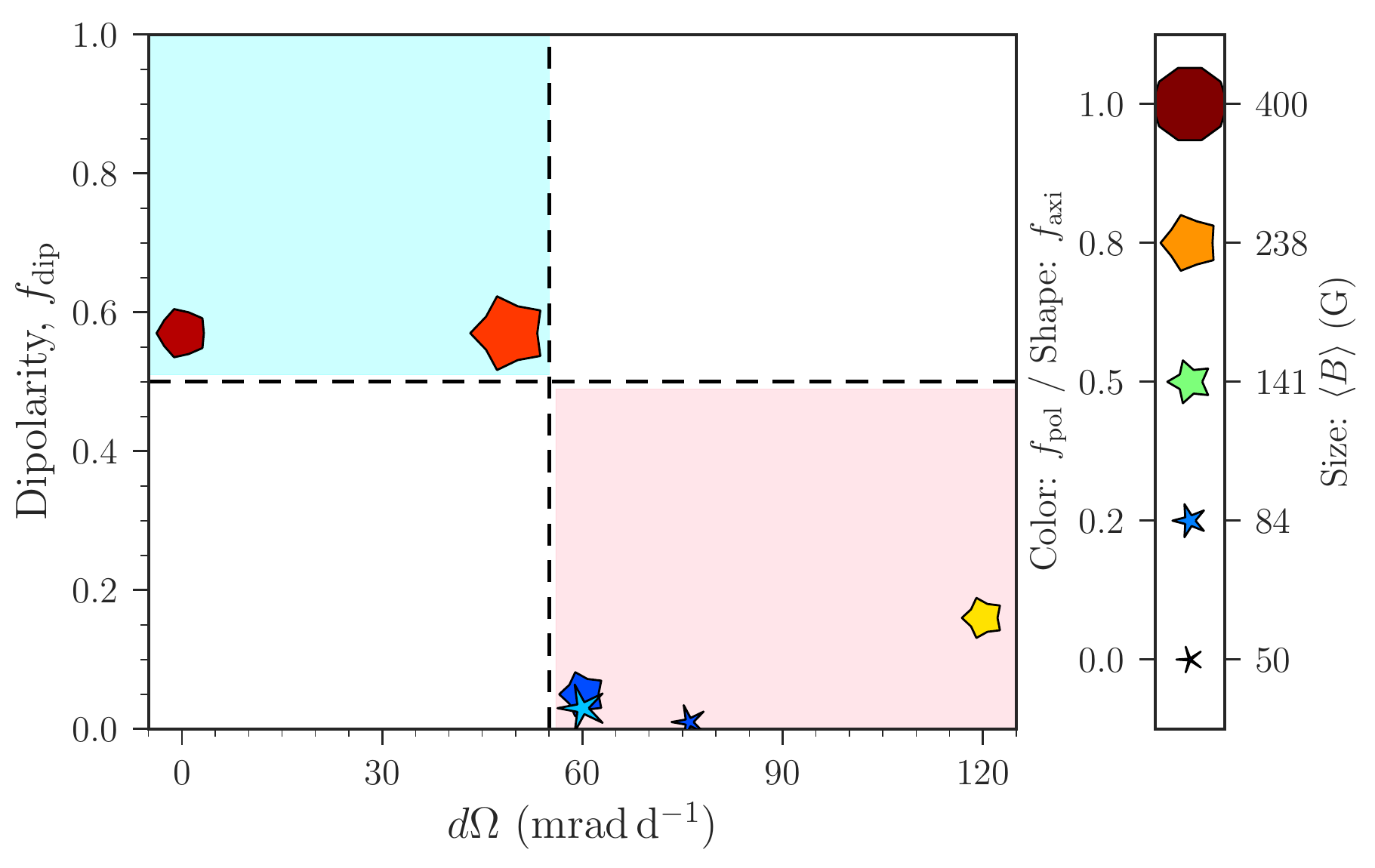}
     \caption{Dipolarity as a function of the surface differential rotation $d\Omega$ measured for a sample of M~dwarfs \citep{DMP08,MDP08}. The surface differential rotation is defined as $d\Omega = \Omega_\mathrm{eq} - \Omega_\mathrm{pol}$, where $\Omega_\mathrm{eq}$ is the angular velocity at the equator and $\Omega_\mathrm{pol}$ at the pole. Symbols are defined as in Figure~\ref{fig:dipOBS}.}
     \label{fig:dipOBSshear}
\end{figure}

Finally, we explored the surface shear achieved in our simulations. We identified that, although quite weak, simulations with multipolar surface magnetic fields favour solar-like differential rotation profiles. In contrast, all dipole dominated simulations yield antisolar differential rotation \citep[similar to][]{A05,DSB06}. Here, we can also draw an observational parallel as surface shears have been measured for some of the stars in Figure~\ref{fig:dipOBS} \citep{DMP08,MDP08}. Because our numerical simulations have a constant rotation period, the transition in the magnetic field morphology with the relative shear reflects the change in the surface shear (Figure~\ref{fig:DRsurf}). Therefore, we use the latitudinal surface shear $d\Omega$ rather than the relative shear as the relevant parameter to consider for observations when the rotation period varies from star to star (from 1 to 9\,d in our sample). Figure~\ref{fig:dipOBSshear} shows the link between the axial dipole contribution to the large-scale magnetic morphology and the measured latitudinal surface shear for M~dwarf stars. The data in Figure~\ref{fig:dipOBSshear} give hints of a sharp transition in the magnetic complexity of M~dwarfs with the increase of $d\Omega$, with strong dipoles preventing significant latitudinal differential rotation at the surface and multipoles co-existing with large latitudinal surface shears. We note that this observational trend also extends to fully convective stars, with those harboring strong dipoles almost rotating as solid bodies, i.e., $d\Omega \sim 0$ \citep{DFC06,MDP08}. However, contrary to the trend in our simulations, we find that the dipole collapses at positive shears for M~dwarfs ($d\Omega \sim 55$\,mrad\,d$^{-1}$). Moreover, none of the stars from \citet{DMP08} or \citet{MDP08} had an antisolar differential rotation \citep[akin to other shear detection in M dwarfs, e.g.,][]{HDD16,ZVC20}. The direct comparison between observations and simulations is thus slightly less straightforward when shear profiles are concerned. It remains therefore to be investigated whether lowering the viscosity and magnetic diffusivity in our simulations can modify the differential rotation profile. For instance, it would be important to test if the antisolar regime found in the present calculations survives in more realistic parameter ranges. Further research is thus necessary to investigate how smaller Ekman numbers and/or larger magnetic Reynolds numbers can impact the transition seen in the differential rotation profile and amplitude.

The parameter space explored in this study offers new insights into the mechanisms controlling the magnetic morphology of stars. Our 3D dynamo simulations show that the magnetic morphology of the large-scale field depends on how much the Lorentz force is able to impact the flow. Although we cannot exclude the possibility that stronger anelastic effects in stars may modify this conclusion, we found that the energy ratio proxy proposed in our work to describe the transition in the magnetic morphology indeed succeeds at describing the varying large-scale magnetic topology of a small sample of M dwarfs featuring similar convective zone geometries, and for which a homogeneous collection of ZDI measurements is available in the literature. This first result leaves room for further numerical explorations aimed at studying the impact of more parameters, such as the size of the convective zone and the rotation rate. These simulations will broaden potential comparisons with stars of different spectral types than the ones considered here, and therefore to further investigate whether the proxy that we propose can be used in a more general context. We also leave for a forthcoming paper the study of whether a radiative interior in the numerical domain is also able to impact the magnetic morphology of the large-scale field and its transition from a mainly dipolar to a mainly multipolar structure, and to modify the conclusions reached here regarding the proxies that best describe where this transition occurs in the parameter space. 

\section*{Acknowledgements}

The authors wish to thank Pascal Petit and Claire Moutou for very fruitful discussions. We also thank the anonymous referees for their insightful comments and suggestions that helped to improve this manuscript. BZ and JFD acknowledge funding from the European Research Council (ERC) under the H2020 research $\&$ innovation  programme (grant agreement $\#740651$ New-Worlds). LJ acknowledges funding by the Institut Universitaire de France. NL acknowledges funding by CNPq, CAPES and FAPEMIG. Numerical simulations were performed using HPC resources from GENCI-CINES (Grants 2020-A0070410970 and 2021-A0090410970) and CALMIP (Grant P19031).

\section*{Data Availability}
The authors confirm that the data supporting the findings of this study are available within the article or through requests to the corresponding author. Numerical simulations were performed with the fluid dynamics code \verb|MagIC| using the open-source library \verb|SHTns|, both publicly available at \url{https://github.com/magic-sph/magic} and \url{https://bitbucket.org/nschaeff/shtns}, respectively.



\bibliographystyle{mnras}
\bibliography{ref} 




\appendix

\section{Kinetic energy length-scale} \label{appendixB}
\begin{figure}
     \centering
     {\includegraphics[width=\columnwidth]{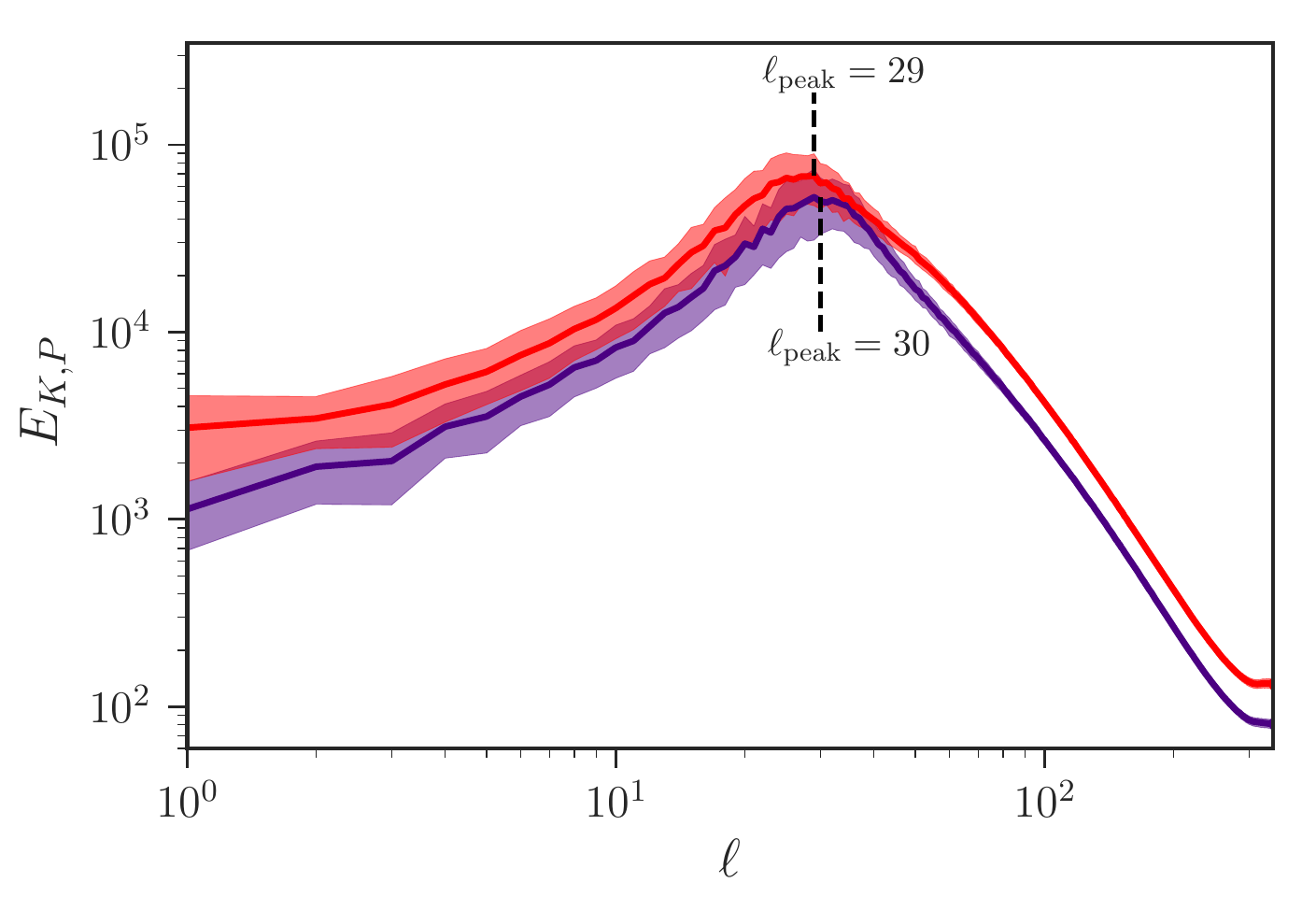}}
     \caption{Time-averaged dimensionless poloidal kinetic energy spectra for the dipolar (red solid line) and multipolar (purple solid line) cases given in Fig.~\ref{fig:FBspec}. Shaded areas correspond to one standard deviation about the time-averaged spectra and the dashed vertical lines mark the location of the peak. }
     \label{fig:ekinspec}
\end{figure}

We compute the dominant scale of convection as the peak of the time-averaged poloidal kinetic energy spectra \citep[][]{SGA19,SGA21}, defined as
\begin{equation} \label{eq:lpeak}
    \ell_\mathrm{peak} = \text{argmax}(E_{K,P}(\ell)).
\end{equation}
Figure~\ref{fig:ekinspec} shows examples of poloidal kinetic energy spectra for one dipolar case (red line) and one multipolar case (purple line). The degree at which the spectra is maximum, $\ell_\mathrm{peak}$, is indicated by a dashed vertical line. These reference dipole and multipole models feature convective flows with similar dominant length scale. Considering the entire set of simulations, we find $\ell_\mathrm{peak}$ ranging from 14 to 45 with a median value of 30.

\section{Averaging strategy} \label{appendix:dipolarity}
\begin{figure}
     \centering
     \includegraphics[width=\columnwidth]{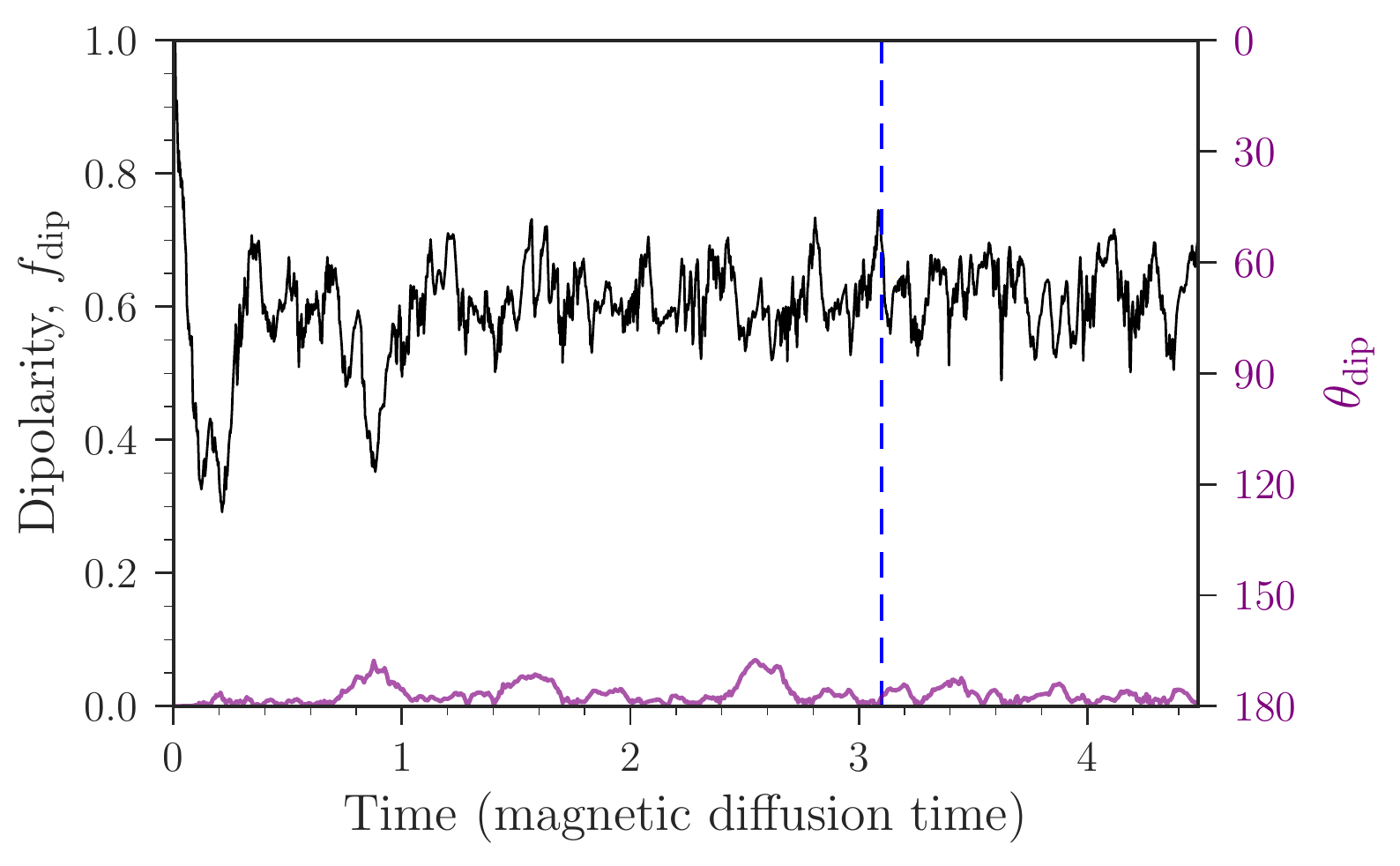}
     \includegraphics[width=\columnwidth]{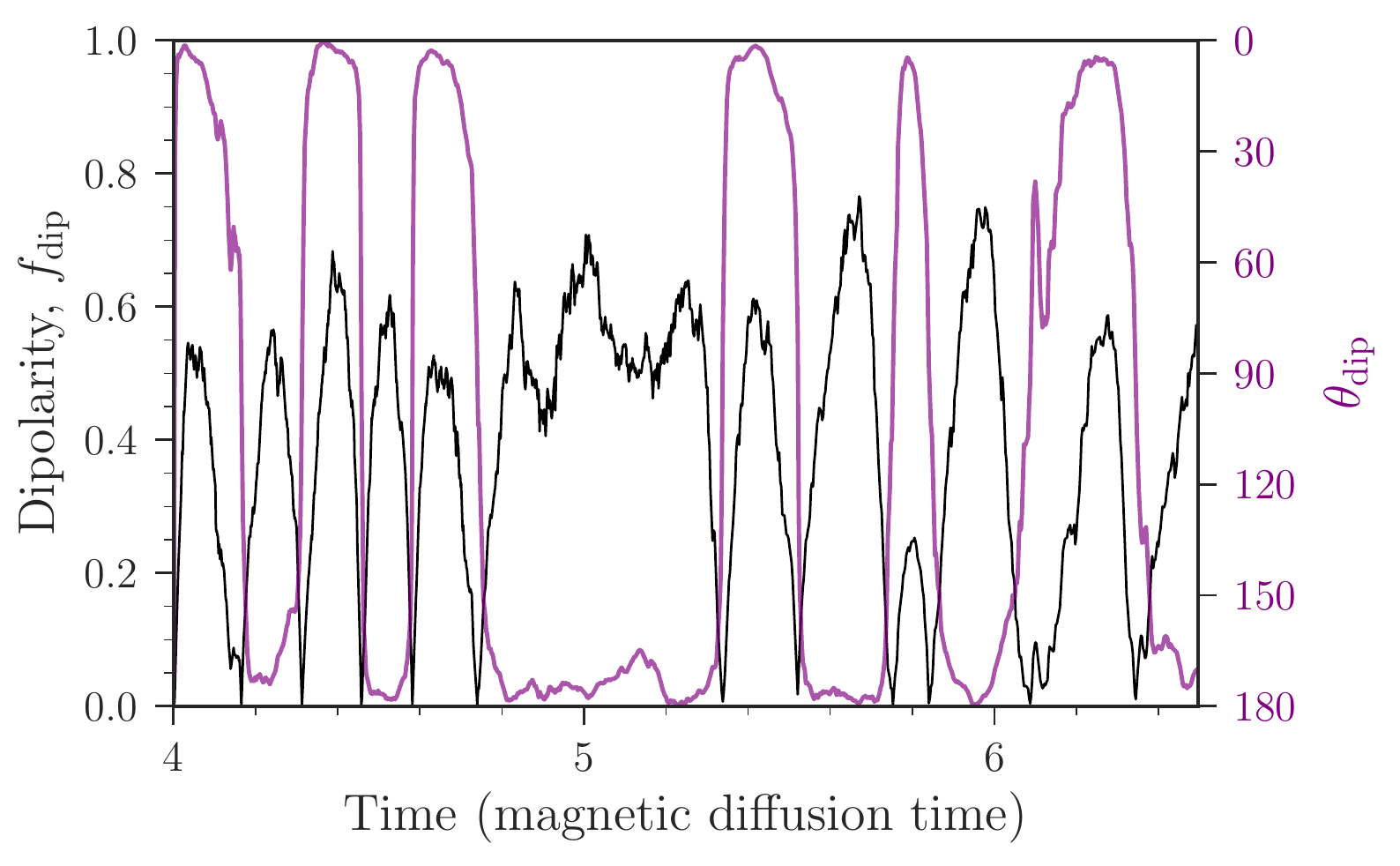}
     \caption{Dipolarity (black line) and tilt angle of the total dipole (purple line) as a function of time (given in units of magnetic diffusion time $\tau_\lambda$). The top plot corresponds to the simulation FC10 and the bottom one to FC11. The vertical blue line indicates the initial time used to compute the time-averaged dipolarity in the top panel. For illustrative purposes only the time-averaged window is shown in the bottom plot.
     }
     \label{fig:tsFdip}
\end{figure}

Figure~\ref{fig:tsFdip} illustrates the time dependence of the dipolarity (Eq.~\ref{eq:fdip}) and the dipole tilt angle ($\theta_\text{dip}$) for two simulations with $\Nrho = 1.5$ in our sample. The simulation FC10 (top panel) shows an axial-dipole that is anti-aligned with the rotation axis ($\theta_\text{dip} \sim 180\degr$) and whose field strength is stable through out the time span of the simulation. For this simulation, we find $\fdip = 0.62 \pm 0.04$ when using an averaging interval $\tau_\mathrm{avg}$ that is defined as the difference between the time at the end of the run ($\tau_\text{end}$) minus a predefined initial time (represented by the blue dashed line in top plot). The bottom panel of Figure~\ref{fig:tsFdip} corresponds to the simulation FC11. The evolution of $\theta_\text{dip}$ evidences a reversing dipole with periodic switches in polarity that occur at irregular intervals of time. We find $\fdip = 0.41 \pm 0.12$ when considering a large number of reversals to compute the time average (achieved after setting $\tau_\mathrm{avg} = 2.5 \tau_\lambda$). 

We remind the reader that all of our simulations were initialized with a dipole of strength $\Lambda = 0.44$ and the solutions we obtained may depend on the initial conditions.

\section{Flow configuration} \label{appendix:flow}
It was proposed in the literature that the dipole collapse is directly linked to a arrangement in the convective flow. Two main quantities characterising the structure of convective flows in the simulations were explored:
\begin{enumerate}
    \item the \textit{columnarity} $\Cwz$ which offers a quantitative way to define columnar flows and is expressed by \begin{equation} \label{eq:columnarity}
\Cwz = \frac{\sum_{s,\phi} \abs{\expval{\va{\omega'}\vdot\ez}_z}}{\sum_{s,\phi} \expval{\abs{\va{\omega'}}}_z} \qc
\end{equation} where $\va{\omega'}$ is the vorticity generated by the non-axisymmetric velocity field \citep{SKA12}. The summation occurs in the equatorial plane and $\avg{\cdot}_z$ represents an average in the axial direction $\ez$;

    \item the relative axial helicity of the flow $\abs{\rHz}$ computed as the average of the absolute contribution from the Northern and Southern hemispheres: $\abs{\rHz} = \qty(\abs{\rHz_{\mathrm{NH}}} + \abs{\rHz_{\mathrm{SH}}})/2$, where each hemispheric contribution is given by \begin{equation} \label{eq:rhelicity}
\rHz_{\mathrm{NH/SH}} = \frac{\expval{\large u_z\omega_z}_{\mathrm{NH/SH}}}{\sqrt{\expval{u_z^2}_{\mathrm{NH/SH}}\expval{\omega_z^2}_{\mathrm{NH/SH}}}}.
\end{equation} 
\end{enumerate}

\begin{figure}
     \centering
     \includegraphics[width=\columnwidth]{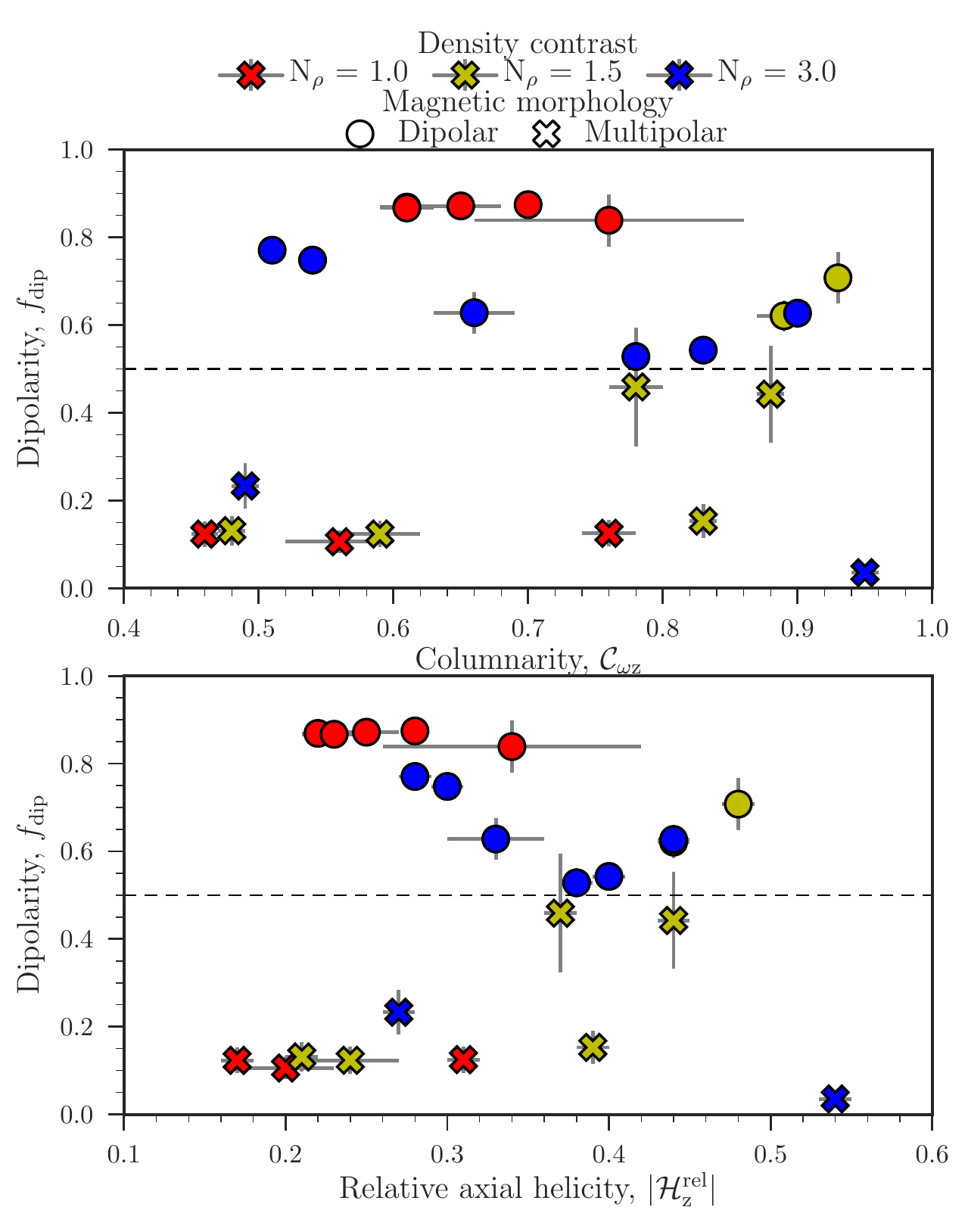}
     \caption{Dipolarity as a function of the flow columnarity (top) and relative axial helicity (bottom). Symbols are defined as in Fig.~\ref{fig:fdip}.
     }
     \label{fig:columnarity}
\end{figure}

The top panel of Fig.~\ref{fig:columnarity} shows $\fdip$ as a function of $\Cwz$ for our data set. The overall result shows a homogeneous distribution of dipole-dominated and complex multipolar surface fields for the explored range of $\Cwz$ (going from 0.4 to 1). It also evidences the lack of correlation between $\fdip$ and $\Cwz$. A possible explanation for this might be the high values of columnarity attained in this work. Prior Boussinesq simulations of \citet{SKA12} found that columnar flows with $\Cwz > 0.5$ can generate either dipolar or multipolar surface magnetic fields, while flows with $\Cwz \lesssim 0.5$ only results in multipolar fields. Indeed if we restrain ourselves to the runs with columnarity around the threshold of $0.5$, we identify three runs FC08, FC15, and FC23, giving hints of a transition to a multipolar branch (all three with $\fdip < 0.25$). Nevertheless, the diversity of magnetic field complexities obtained at high–$\Cwz$ makes the columnarity a poor proxy to describe the dipolar collapse.

Often associated with the magnetic field amplification in the dynamo framework (through the so-called $\alpha$–effect), the decrease in the flow's relative axial helicity has also been suggested to cause the dipole breakdown \citep{SKA12}. The bottom panel of Fig.~\ref{fig:columnarity} shows the dependency of $\abs{\rHz}$ with the different magnetic morphologies. The simulations yield weak to moderate relative helicity values, $\abs{\rHz} < 0.6$, that are consistent with the values obtained in previous works \citep{T14,GOD17}. It is apparent from Fig.~\ref{fig:columnarity} that the only case displaying $\fdip \approx 0$ features the highest helicity in our sample. On the other hand, the strongest dipoles possess weak helicity values with $\abs{\rHz}$ spread around $0.28$ (corresponding to five dipolar dynamos obtained for $\Nrho = 1.0$ and the two strongest dipoles for $\Nrho = 3.0$). These results suggest that the magnetic morphology is unaffected by $\abs{\rHz}$ for the parameter space we explored. Although these findings differ from some published studies \citep[e.g.,][]{SKA12}, they are consistent with mean-field simulations of \citet{LHT07} and the 3D simulations of \citet{B08} mimicking the interior of a fully convective M~dwarf. The likely cause for these differences is that the mean-helicity becomes a poor approximation for the $\alpha$–effect in some cases \citep{SRS07,WRT18}.

These results corroborate earlier suggestions of \citet{GOD17}, who argued that hydrodynamic transitions in the flow (e.g. measured by $\Cwz$ or $\rHz$) would only capture the dipole collapse in systems where the Lorentz force plays a minor role in the flow dynamics.


\bsp	
\label{lastpage}
\end{document}